\documentclass[11pt]{article}

\def\mytitle#1{\setcounter{equation}{0}
\setcounter{footnote}{0}
\begin{flushleft}\Large\textbf{#1}\end{flushleft}
\vspace{0.25cm}}
\def\myname#1{\leftline{{\large #1}}\vspace{-0.13cm}}
\def\myplace#1#2{\small\begin{flushleft}\textit{#1}\\
\texttt{#2}\end{flushleft}}

\usepackage{graphicx}
\usepackage{amsmath, amssymb, graphics}
\begin{document}

\mytitle{Red-shift parametrizations of dark energy and
observational constraint on their parameters: Galileon gravity as
background}

\vskip0.2cm \myname{Prabir Rudra\footnote{prudra.math@gmail.com}}
\myplace{Department of Mathematics, Asutosh College, Kolkata-700
026, India.}{}

\vskip0.1cm \myname{Chayan
Ranjit\footnote{chayanranjit@gmail.com}} \myplace{Department of
Mathematics, Egra S. S. B. College, Purba Medinipur-721429, W.B.
India.}{} \vskip0.1cm

\vskip0.1cm \myname{Sujata
Kundu\footnote{sujatakundu10@gmail.com}} \myplace{Department of
Information Technology, Narula Institute of Technology,
Kolkata-700109,India}{} \vskip0.1cm

\begin{abstract}
In this work, FRW universe filled with dark matter (perfect fluid
with negligible pressure) along with dark energy in the background
of Galileon gravity is considered. Four dark energy models with
different EoS parametrizations have been employed namely, Linear,
CPL, JBP and Logarithmic parametrizations. From Stern, Stern+BAO
and Stern+BAO+CMB joint data analysis, we have obtained the bounds
of the arbitrary parameters $\omega_{0}$ and $\omega_{1}$ by
minimizing the $\chi^{2}$ test. The best-fit values and bounds of
the parameters are obtained at 66\%, 90\% and 99\% confidence
levels which are shown by closed confidence contours in the
figures. For the logarithmic model unbounded confidence contours
are obtained and hence the model parameters could not be finitely
constrained. The distance modulus $\mu$(z) against redshift $z$
has also been plotted for our predicted theoretical models for the
best fit values of the parameters and compared with the observed
Union2 data sample and SNe Type Ia 292 data and we have shown that
our predicted theoretical models permits the observational data
sets. From the data fitting it is seen that at lower redshifts
$(z<0.3)$ the SNe Type Ia 292 data gives a better fit with our
theoretical models compared to the Union2 data sample. So, from
the data analysis, SNe Type Ia 292 data is the more favoured data
sample over its counterpart given the present choice of free
parameters. From the study, it is also seen that the logarithmic
parametrization model is less supported by the observational data.
Finally we have generated the plot for the deceleration parameter
against the redshift parameter for all the theoretical models and
compared the results with the work of Farooq et al, 2013.
\end{abstract}

\vspace{5mm}


\vspace{5mm}

{\it Pacs. No.: 98.80.-k, 98.80.Es, 95.35.+d, 95.36.+x}\\

\vspace{5mm}

\section{Introduction}
Over the past decade the most pressing goal for modern cosmology
has been to find a proper explanation for the recent cosmic
acceleration as confirmed from the observations of type Ia
Supernovae and Cosmic Microwave Background (CMB)
\cite{Perlmutter,Riess,Riess1,Bennet,Sperge} radiation. The
standard explanation invokes an unknown component which has the
property of positive energy density and negative pressure, known
as ``dark energy''. Different observations have also been shown
that the contributions of dark energy and dark matter are
respectively about $70\%$ and $26\%$ of the total energy of the
universe. The phenomenon of the accelerated expansion of the
universe has also been strongly confirmed by some other
independent experiments like Sloan Digital Sky Survey (SDSS)
\cite{Adel}, Baryonic Acoustic Oscillation (BAO)
\cite{Eisenstein}, WMAP data analysis \cite{Briddle,Spergel} etc.
Over the past decade, there have been many theoretical models
which mimic the dark energy behaviors. The simplest of them is the
cosmological constant in which the equation of state is
independent of the cosmic time and which fit the observations
well. This model is the so-called $\Lambda$CDM model, containing a
mixture of cosmological constant $\Lambda$ and cold dark matter
(CDM). However, this model is plagued by two famous problems,
namely ``fine-tuning'' and the ``cosmic coincidence'' problem. In
order to solve these two problems, numerous dynamical dark energy
models were suggested, whose equation of state evolves with cosmic
time. The scalar field or quintessence \cite{Peebles,Cald} is one
of the most favourite candidate of dark energy which produce
sufficient negative pressure to drive the cosmic acceleration. In
order to alleviate the cosmological-constant problems and explain
the accelerated expansion, many dynamical dark energy models have
been proposed, such as K-essence, Tachyon, Phantom, quintom,
Chaplygin gas model, etc \cite{Arme,Sen,Cald1,Feng,Kamen}.
Interaction between dark energy and dark matter have been
identified as a very important tool to address these problems. As
a result interacting dark energy models including Modified
Chaplygin gas \cite{Debnath}, holographic dark energy model
\cite{Cohen} and braneworld model \cite{Sahni} have been proposed.
Since the discovery of General theory of relativity, the most
pressing objective for cosmology has been found as the complete
theory of quantum gravity i.e., the theory of everything.
Relentless efforts and their results can be found widely in
literature. Recently, such efforts using the principle of quantum
gravity, led to the proposals of agegraphic dark energy (ADE) and
the new agegraphic dark energy (NADE) models by Cai \cite{Cai} and
Wei et al. \cite{Wei} respectively. Many theoretical models have
been tallied with the observations using different data sets, say
TONRY, Gold sample data sets \cite{Paddy1,Riess1,Tonry,Barris}
etc. In Einstein's gravity, the modified Chaplygin gas
\cite{Debnath} best fits with the 3 year WMAP and the SDSS data
with the choice of parameters $A =0.085$ and $\alpha = 1.724$
\cite{Lu} which are improvements over the previous ones $-0.35 < A
< 0.025$ \cite{Jun}.\\

Another possibility of the accelerated expansion is attributed to
the fact that general relativity is only accurate on small scales
and therefore needs modification on cosmological distances. This
led to the proposal of modified gravity theories. In this case,
cosmic acceleration would arise not to from dark energy as a
constituent substance but rather from the dynamics of modified
gravity. Modified gravity constitutes an interesting dynamical
alternative to $\Lambda$CDM cosmology regarding given the fact
that it is also able to describe the current acceleration in the
expansion of our universe. One of the simplest modified gravity is
DGP brane-world model \cite{Dvali}. The other alternative approach
dealing with the acceleration problem of the Universe is changing
the gravity law through the modification of action of gravity by
means of using $f(R)$ gravity \cite{An,Noj0} instead of the
Einstein-Hilbert action. Some of these models, such as $1/R$ and
logarithmic models, produced an accelerated expansion for the
Universe at the present time \cite{clif}. Other modified gravity
includes $f(T)$ gravity, $f(G)$ gravity, Gauss-Bonnet gravity,
Horava-Lifshitz gravity, Brans-Dicke gravity, etc
\cite{Yer,Noj,An1,Hora,Brans}.\\

Of late an infrared modification of classical gravitation was
proposed, which is a generalization of the 4D effective theory in
the DGP model \cite{Nicolis1}. The theory considers a
self-interaction term of the form $~^{\fbox{}}~\phi\left(\nabla
\phi\right)^{2}$ in order to recover GR in high density regimes.
The most notable feature of the theory is that it is invariant
under the Galileon shift symmetry $\delta_{\mu}\phi$ $\rightarrow$
$\delta_{\mu}\phi+c_{\mu}$ in the Minkowski background. Because of
this invariance, the equation of motion remains a second order
differential equation, preventing the introduction of extra
degrees of freedom, which are usually associated with
instabilities. So we assume the FRW universe in Galileon gravity
model filled with the dark matter and dark energy.\\

As stated earlier, in order to explain the evolution of the
universe, various dark energy models have been proposed. All these
models must be constrained by astronomical observations in order
to bring about a consistency between the proposed theoretical
models and the observational data. For any dark energy model, the
Equation of state (EoS) parameter $\omega$ plays a vital role and
can reveal the nature of the model which is responsible for the
cosmic acceleration. Different form of EoS lead to different
dynamical nature and may influence the evolution of the Universe
considerably. The EoS parameter $\omega$ and its time derivative
with respect to Hubble time are currently constrained by the
distance measurements of the type Ia supernova. From literature it
is seen that the current observational data constrain the range of
EoS as $-1.38 < \omega <-0.82$ \cite{Melchiorri1}. Recently, the
combination of WMAP3 and Supernova Legacy Survey data showed a
significant constraint on the EOS $\omega =-0.97^{+0.07}_{-0.09}$
for the dark energy in a flat universe \cite{Seljak1}. Recently,
some parametrizations for the variation of EOS parameters
$\omega(z)$ have been proposed describing the DE component. We
will discuss them in detail in section 3.\\

It is a known fact that the standard big bang model of cosmology
is crippled by various problems such as the horizon problem, the
flatness problem, the singularity problem, etc. Considering the
inflationary model, we have been able to explain the flatness
problem. Now for a flat universe filled with matter and dark
energy, an accurate knowledge of $\Omega_{de}$ and $\omega_{de}$
for dark energy can only be obtained if we have a proper knowledge
of $\Omega_{m}$ for dark matter and $H(z)$ \cite{Paddy1, Paddy2}.
For $z>0.01$, we see that the TONRY data with 230 data points
\cite{Tonry} with 23 data points from Barris et al. \cite{Barris}
is quite acceptable. In the redshift range $1<z<1.6$, the GOLD
sample of Riess et al. \cite{Riess1} with 156 data points is
valid. From the CMBR data one finds that
$\Omega_{\Lambda}+\Omega_{m}=1$ \cite{Spergel}. From the most
recently obtained Riess data set the best fit value of
$\Omega_{m}$ is obtained as $0.31\pm 0.04$. This value matches
with the value of $\Omega_{m}=0.29^{+0.05}_{-0.03}$ obtained from
the previous Riess data \cite{Riess}. In Chowdhury et al.
\cite{Paddy1} the value was found to be $0.31\pm 0.08$. For a flat
universe, the best fit value for the equation of state parameter
$w$ for union 2 data sample is
$\omega=-0.997^{+0.050}_{-0.054}$(stat)$^{+0.077}_{-0.082}$(stat+sys
together). For a curved one, the value is
$\omega=-1.038^{+0.056}_{-0.059}$(stat)$^{+0.093}_{-0.097}$(stat+sys
together) \cite{Amanullah1}.\\

The success of any dark energy or modified gravity model, depends
basically on its consistency with the observational data. This is
our basic motivation for the work. We reconstruct the hubble
parameter $H$ using the parameters of dark energy, dark matter and
modified gravity. Then we set up a comparison scenario between the
reconstructed $H$ ($H_{theoretical}$) and the values of $H$
obtained from observational data ($H_{observational}$). This is
accomplished by the procedure of chi-square test. Recently,
numerous works can have appeared in the literature, which aims at
constraining dark energy model parameters using Stern and union2
data sets \cite{Ranjit1, Ranjit2, Chakraborty1, Paul1, Paul2}. The
basic concepts of Galileon gravity theory are presented in section
2. The behaviour of some reconstructed cosmological parameters is
studied in section 3. Some basic calculations using the Galileon
gravity and parametrized dark energy models is given in section 4.
The observational data analysis tools in observed Hubble data
(OHD) or $H(z)$-$z$ (Stern), OHD+BAO and OHD+BAO+CMB for
$\chi^{2}$ minimum test will be studied in section 5 and we will
also investigate the bounds of unknown parameters
$(\omega_{0},\omega_{1})$ for the various dark energy
parametrizations by fixing other parameters. The best-fit values
of the parameters are obtained at 66\%, 90\% and 99\% confidence
levels. The redshift magnitude observations for our theoretical
parametrization models in Galileon gravity for the best fit values
of the parameters and the observed SNe Ia union2 data sample is
studied in section 6. Analysis of our theoretical models with SNe
Type Ia 292 data is given in section 7. Finally the paper ends
with a discussion in section 8.

\section{Basic equations of Galileon gravity}

The Galileon gravity theory is described by the action
\cite{Nicolis1,Silva1, Deffayet1,Deffayet2,Chow1}:
\begin{equation}\label{Lag}
S=\int d^{4} x \sqrt{-g}\left[\phi R- \frac{w}{\phi} \left(\nabla
\phi\right)^{2}+ f(\phi)^{\fbox{}}~\phi \left(\nabla
\phi\right)^{2}+{\cal L}_{m}\right]
\end{equation}
where $\phi$ is the Galileon field, \textit{w} is known as
Brans-Dicke (BD) parameter, the coupling function $f(\phi)$ has
dimension of length, $\left(\nabla
\phi\right)^{2}=g^{\mu\nu}\nabla_{\mu}\phi \nabla_{\nu}\phi$,~
$^{\fbox{}}~\phi=g^{\mu\nu}\nabla_{\mu}\nabla_{\nu}\phi$ and
${\cal L}_{m}$ is the matter Lagrangian. Variation of Eqn. (1)
with respect to the metric $g_{\mu\nu}$ gives the Einstein's
equations,

$$G_{\mu
\nu}=\frac{T_{\mu\nu}}{2\phi}+\frac{1}{\phi}\left(\nabla_{\mu}\nabla_{\nu}\phi-g_{\mu\nu}~^{\fbox{}}~\phi\right)
+\frac{w}{\phi^{2}}\left[\nabla_{\mu}\phi\nabla_{\nu}\phi-\frac{1}{2}g_{\mu\nu}\left(\nabla\phi\right)^{2}\right]$$
\begin{equation}
-\frac{1}{\phi}\left\{\frac{1}{2}g_{\mu\nu}\nabla_{\lambda}[f(\phi)\left(\nabla
\phi\right)^{2}]\nabla^{\lambda}\phi-\nabla_{\mu}[f(\phi)\left(\nabla
\phi\right)^{2}]\nabla_{\nu}\phi+f(\phi)\nabla_{\mu}\phi\nabla_{\nu}\phi~^{\fbox{}}~\phi\right\}
\end{equation}
For the Friedmann-Robertson-Walker background metric, the
Einstein's field eqns for Galileon gravity gives,
\begin{equation}
3H^{2}=\frac{\rho}{2\phi}-3HI+\frac{w}{2}I^{2}+\phi^{2}f(\phi)\left(3H-\frac{\alpha_{1}}{2}I\right)I^{3}
\end{equation}
and
\begin{equation}
-3H^{2}-2\dot{H}=\frac{p}{2\phi}+\dot{I}+I^{2}+2HI+\frac{w}{2}I^{2}-\phi^{2}f(\phi)\left(\dot{I}+\frac{2+\alpha_{1}}{2}I^{2}\right)I^{2}
\end{equation}
where $H(t)=\frac{\dot{a}}{a}$, $I(t)=\frac{\dot{\phi}}{\phi}$ and
$\alpha_{n}[\phi(t)]=\frac{d^{n}\ln f}{d \ln \phi^{n}}$ .
\vspace{5mm}

Here $\rho=\rho_{de}+\rho_{m}$ and $p=p_{de}+p_{m}$, where
$\rho_{m}$, $p_{m}$ are the energy density and pressure of the
dark matter and $\rho_{de}$, $p_{de}$ are respectively the energy
density and pressure contribution of some dark energy.

\section{Some parametrizations on EoS parameter}
Recently some parametrizations for the variation of EOS parameters
$\omega(z)$ have been proposed describing the DE component. Here
we discuss some of them.

\subsection{Linear Parametrization}
Here the EoS is given by \cite{Cooray1}

\begin{equation}
\omega(z)=\omega_{0}+\omega_{1}z
\end{equation}
Here $\omega_{0}=-1/3$ and $\omega_{1} =-0.9$ with $z<1$ when
Einstein gravity has been considered. This grows increasingly
unsuitable for $z>1$.
\subsection{Chevallier-Polarski-Lindler (CPL) Parametrization}
Here the EoS is given by
\begin{equation}
\omega(z)=\omega_{0}+\omega_{1}\frac{z}{1+z}
\end{equation}
This ansatz was first discussed by Chevallier and Polarski
\cite{Chevallier1} and later studied more elaborately by Linder
\cite{Linder1}. In Einstein gravity the best fit values for this
model while fitting with SN1a gold data set are $\omega_{0}=-1.58$
and $\omega_{1}= 3.29$. This parametrization has several
advantages over its predecessor. Some of them are: 1) It has a
manageable 2-dimensional phase space; 2) reduces to the old linear
redshift behaviour at low redshift; 3) has a well behaved and
bounded behaviour for high redshift; 4) highly accurate in
reconstructing many scalar field equations of state and the
resulting distance-redshift relations; 5) highly sensitive to
observational data and 6) simple physical interpretation. In
addition to the above advantages, the new parametrization is also
more accurate than the previous one. The most remarkable thing
about the parametrization is that, it reconstructs the
distance-redshift behaviour of the SUGRA model \cite{Brax1} to
0.2\% over the entire range out to the last scattering surface
$(z\approx 1100)$.

\subsection{Jassal-Bagla-Padmanabhan (JBP) Parametrization}
Here the EoS is given by \cite{Jassal1}
\begin{equation}
\omega(z)=\omega_{0}+\omega_{1}\frac{z}{(1+z)^2}
\end{equation}
A fairly rapid evolution of this EoS is allowed so that
$\omega(z)\geq -1/2$ at $z>0.5$ is consistent with the supernovae
observation in Einstein gravity. From the EoS it can be seen that
$\omega(0)=\omega_{0}$ and $\omega(\infty)=\omega_{0}$. For a
standard cosmological model with a hot big bang to be valid, we
must have $\omega(z\gg 1)\leq -1/3$.

\subsection{Logarithmic Parametrization}
Here the EoS is given by \cite{Efs1, Silva2}
\begin{equation}
\omega(z)=\omega_{0}+\omega_{1}\log(1+z)
\end{equation}
This model is sparsely found in literature, and therefore very
little is known about the behaviour of this model. It has been
used now and then for parametrization of dark energy. Here we will
analyze it and try to compare it with other models. This will help
us to understand its importance in cosmology and also to find out
the reason why it has been so seldom studied in past. This is the
basic reason for the inclusion of this model in the present study.

\section{Basic calculations on Galileon gravity using various EoS parametrizations}
Here we will consider the flat, homogeneous and isotropic universe
described by the FRW background metric. The corresponding Einstein
equations in Galileon gravity are given by eqns.(3) and (4). The
DE equation of state is taken as $p_{de}=\omega(z)\rho_{de}$. We
neglect the pressure contribution of matter, i.e. $p_{m}=0$. The
independent conservation equations for DE and DM are given as,

\begin{equation}
\dot{\rho_{de}}+3H\left(\rho_{de}+p_{de}\right)=0
\end{equation}
and
\begin{equation}
\dot{\rho_{m}}+3H\rho_{m}=0
\end{equation}
From eqn.(10), we have
\begin{equation}
\rho_{m}=\rho_{m0}\left(1+z\right)^{3}
\end{equation}
where we have used the relation $a=\frac{1}{1+z}$, $z$ being the
cosmological redshift parameter. Here $\rho_{m0}$ is a constant.
Integrating equation (9) for the four DE models we get four
different solutions for DE density. Now, we consider dimensionless
density parameters $\Omega_{m0}=\frac{\rho_{m0}}{3H_{0}^{2}}$ and
$\Omega_{de0}=\frac{\rho_{de0}}{3H_{0}^{2}}$ and for simplicity
choosing $f(\phi)=f_{0}\phi^{n}$ and $\phi=\phi_{0}a^{m}$
($f_{0}>0,~\phi_{0}>0,~m>0,~n>0$) in the power law form,
\cite{Ranjit2} we have the expression for Hubble parameter $H$ in
terms of redshift parameter $z$ as follows (from eq. 3):
$$H(z)=\frac{1}{2 \left(6 f_{0} m^3 \phi_{0}^{3+n}-f_{0} m^4 n \phi_{0}^{3+n}\right)}\left[-(m^2 w-6m-6) (1+z)^{m (2+n)}
\phi_{0}\right.~~~~~~~~~~~~~~~~~~~~~~~~~~~~~~~~~~~~~~~~~~~~~~~~~~~~~~~~~~~~~~~~~~~~$$
$$\left.+ \left\{\left((m^{2} w-6m-6)(1+z)^{m (2+n)} \phi_
{0}\right)^2-f_{0} m^4 n \phi_ {0}^{3+n}-12H_{0}^{2}
\left((1+z)^{3+m (3+n)} \Omega_ {m_{0}}+\right.\right.\right.$$
\begin{equation}
\left.\left.\left. (1+z)^{m (3+n)}
\frac{\rho_{de}}{3H_{0}^{2}}\right) 6 f_{0} m^3 \phi_
{0}^{3+n}\right\}^{\frac{1}{2}}\right]
\end{equation}

\subsection{Expressions for Linear parametrization}
Using equation (5), we integrate equation (9) and get the
expression for DE density as
\begin{equation}
\rho_{de}=\rho_{de0}\left(1+z\right)^{3\left(1+\omega_{0}-\omega_{1}\right)}exp\left\{3\omega_{1}\left(1+z\right)\right\}
\end{equation}
where $\rho_{de0}$ is a constant. In terms of the above mention
dimensionless density parameters and redshift parameter (z) we
calculate the Hubble parameter as follows,

$$H_{lin}(z)=\left[\frac{3\left(1+z\right)^{m(2+n)+3\omega_{1}}\phi_{0}\left(1+m-\frac{1}{3}m^{2}w\right)}
{f_{0}m^{3}\phi_{0}^{3+n}\left(1+z\right)^{3\omega_{1}}\left(6-mn\right)}+\left(\left(\left(36\left(1+z\right)^{2m(2+n)+6\omega_{1}}\phi_{0}^{2}\left(1+m^{2}-
\frac{1}{6}m^{2}w\right)^{2}\right)\right.\right.\right.$$

$$\left.\left.\left.+12H_{0}^{2}\left(\left(1+z\right)^{m+m(2+n)+3\omega_{1}}\Omega_{m0}
\left(1+3z+3z^{2}+z^{3}\right)+exp\left\{3\omega_{1}\left(1+z\right)\right\}\left(1+z\right)^{m+m(2+n)
+3\omega_{0}}\Omega_{de0}\right.\right.\right.\right.$$

\begin{equation}
\left.\left.\left.\left.\left(1+3z+3z^{2}+z^{3}\right)\right)f_{0}m^{3}\left(1+z\right)^{3\omega_{1}}\phi_{0}^{3+n}
\left(mn-6\right)\right)^{1/2}\right)\times
\frac{1}{2f_{0}m^{3}\left(1+z\right)^{3\omega_{1}}\phi_{0}^{3+n}
\left(mn-6\right)}\right]^{1/2}
\end{equation}

\subsection{Expressions for CPL parametrization}
For this DE parametrization the DE density is obtained as,
\begin{equation}
\rho_{de}=\rho_{de0}\left(1+z\right)^{3\left(1+\omega_{0}+\omega_{1}\right)}exp\left\{3\omega_{1}\left(1+z\right)\right\}
\end{equation}
Using the dimensionless density parameters, the expression for the
Hubble parameter in terms of the redshift is obtained as,
$$H_{CPL}(z)=\left[\frac{3\left(1+z\right)^{m(2+n)}\phi_{0}\left(1+m-\frac{1}{6}m^{2}w\right)}
{f_{0}m^{3}\phi_{0}^{3+n}\left(6-mn\right)}+\frac{1}{2f_{0}m^{3}\phi_{0}^{3+n}\left(6-mn\right)}
\left(\left(\left(36\left(1+z\right)^{2m(2+n)}\phi_{0}^{2}\left(1+m-
\frac{1}{6}m^{2}w\right)^{2}\right)\right.\right.\right.$$

$$\left.\left.\left.+12H_{0}^{2}\left(\left(1+z\right)^{m+m(2+n)}\Omega_{m0}
\left(1+3z+3z^{2}+z^{3}\right)+exp\left\{\frac{3\omega_{1}}{(1+z)}\right\}\left(1+z\right)^{m+m(2+n)
+3\omega_{0}+3\omega_{1}}\Omega_{de0}\right.\right.\right.\right.$$

\begin{equation}
\left.\left.\left.\left.\left(1+3z+3z^{2}+z^{3}\right)\right)f_{0}m^{3}\left(1+z\right)^{3\omega_{1}}\phi_{0}^{3+n}
\left(mn-6\right)\right)^{1/2}\right)\times
\frac{1}{2f_{0}m^{3}\phi_{0}^{3+n} \left(mn-6\right)}\right]^{1/2}
\end{equation}

\subsection{Expressions for JBP parametrization}
Here the DE density is obtained as,
\begin{equation}
\rho_{de}=\rho_{de0}\left(1+z\right)^{3\left(1+\omega_{0}\right)}exp\left\{-\frac{3\omega_{1}\left(z+\frac{1}{2}\right)}{\left(1+z\right)^{2}}\right\}
\end{equation}
Using the dimensionless density parameters, the expression for the
Hubble parameter in terms of the redshift is obtained as,

$$H_{JBP}(z)=\frac{1}{\sqrt{2}}\left[\left(\frac{1}{f_{0}
m^{3} (-6+m n)}exp\left\{-\frac{3(\omega_{1}+2
z\omega_{1})}{2(1+z)^2}\right\}\phi_{0}^{-3-n}\right.\right.\\
\left(exp\left\{\frac{3 (\omega_{1}+2 z \omega_{1})}{2
(1+z)^{2}}\right\} (-6+m (-6+m w))\right.$$

$$(1+z)^{m
(2+n)}\phi_{0}\left(exp\left\{\frac{3 (\omega_{1}+2 z
\omega_{1})}{(1+z)^{2}}\right\}(6+m(6-m
w))^{2}(1+z)^{2m(2+n)}\phi_{0}^2+\right.
12e^{\frac{3(\omega_{1}+2z\omega_{1})}{2(1+z)^{2}}}f_{0} H_{0}^{2}
m^{3}(-6+mn)
\\$$

\begin{equation}
\left.\left.\left.\left.(1+z)^{3+m(3+n)}\phi_{0}^{3+n}\left(exp\left\{\frac{3
(\omega_{1}+2 z \omega_{1})}{2(1+z)^2}\right\}
\Omega_{m0}+(1+z)^{3\omega_{0}}\Omega_{x0}\right)\right)^{1/2}\right)\right)^{1/2}\right]
\end{equation}

\subsection{Expressions for Logarithmic parametrization}
For this parametrization model, the DE density is obtained as,
\begin{equation}
\rho_{de}=\rho_{de0}\left(1+z\right)^{3\left(1+\omega_{0}+\omega_{1}\right)}
\end{equation}
Using the dimensionless density parameters, the expression for the
Hubble parameter in terms of the redshift is obtained as,

$$H_{Log}(z)=\left[\left(\frac{1}{2f_{0} m^{3}\left(-6+mn\right)}
\phi_{0}^{-3-n}\left(-6(1+z)^{m(2+n)}\phi_{0}-6m(1+z)^{m (2+n)}
\phi_{0}+m^{2}w\left(1+z\right)^{m(2+n)}\phi_{0}+\left(\left(6+6m\right.\right.\right.\right.\right.$$
\begin{equation}
\left.\left.\left.\left.\left.-m^2w\right)^2(1+z)^{2m(2+n)}
\phi_{0}^2+12f_{0}H_{0}^{2} m^{3} \left(-6+mn\right)
\left(1+z\right)^{3+m(3+n)}\phi_{0}^{3+n}
\left(\Omega_{m0}+\left(1+z\right)^{3
\left(\omega_{0}+\omega_{1}\right)}
\Omega_{de0}\right)\right)^{1/2}\right)\right)\right]^{1/2}
\end{equation}

\section{Observational data analysis mechanism}
In the following subsections, we present the observational data
analysis mechanism for Stern, Stern+BAO and Stern+BAO+CMB
observations. We use the $\chi^{2}$ minimum test from theoretical
Hubble parameter with the observed data set and find the best fit
values of unknown parameters for different confidence levels. From
the above expressions, we see that $H(z)$ contains the unknown
parameters like $\omega_{0},~\omega_{1}$, $\Omega_{m0}$,
$\Omega_{de0}$, $w$, $n$, $m$, $f_0$, $\phi_0$. Now the relation
between two parameters will be obtained by fixing the other
parameters and by using observational data set. Eventually the
bounds of the parameters will be obtained by using this
observational data analysis mechanism.
\[
\begin{tabular}{|c|c|c|}
 \hline
  ~~~~~~$z$ ~~~~& ~~~~$H(z)$ ~~~~~& ~~~~$\sigma(z)$~~~~\\
  \hline
  0 & 73 & $\pm$ 8 \\
  0.1 & 69 & $\pm$ 12 \\
  0.17 & 83 & $\pm$ 8 \\
  0.27 & 77 & $\pm$ 14 \\
  0.4 & 95 & $\pm$ 17.4\\
  0.48& 90 & $\pm$ 60 \\
  0.88 & 97 & $\pm$ 40.4 \\
  0.9 & 117 & $\pm$ 23 \\
  1.3 & 168 & $\pm$ 17.4\\
  1.43 & 177 & $\pm$ 18.2 \\
  1.53 & 140 & $\pm$ 14\\
  1.75 & 202 & $\pm$ 40.4 \\ \hline
\end{tabular}
\]
{\bf Table 1:} The Hubble parameter $H(z)$ and the standard error
$\sigma(z)$ for different values of redshift $z$.

\subsection{Analysis with Stern data}

We analyze the model, using observed value of Hubble parameter at
different redshifts (twelve data points) listed in observed Hubble
data by Stern et al. \cite{Stern1}. The Hubble parameter $H(z)$
and the standard error $\sigma(z)$ for different values of
redshift $z$ are given in Table 1. Since we are using testing of
hypothesis, so before proceeding, we form our null and alternate
hypothesis which are given below.\\

\textbf{Null Hypothesis: ~~~~~~~~~~~~~~~~~~~~$H_{0}$: $H_{theoretical}=H_{observational}$}\\

\textbf{Alternative Hypothesis:~~~~~~~~~~~~$H_{1}$:
$H_{theoretical}\neq H_{observational}$}\\\\
Here we test the null hypothesis $H_{0}$ against the alternative
hypothesis $H_{1}$. For this purpose we first form the $\chi^{2}$
statistics as a sum of standard normal distribution as follows:
\begin{equation}
{\chi}_{Stern}^{2}=\sum\frac{(H(z)-H_{obs}(z))^{2}}{\sigma^{2}(z)}
\end{equation}
where $H(z)$ and $H_{obs}(z)$ are theoretical and observational
values of Hubble parameter at different redshifts respectively.
Here, $H_{obs}$ is a nuisance parameter and can be safely
marginalized. We consider the present value of Hubble parameter
$H_{0}$ = 72 $\pm$ 8 Kms$^{-1}$ Mpc$^{-1}$ and a fixed prior
distribution. Here we shall determine best fit value of the
parameters ($\omega_{1},\omega_{0}$) by minimizing the above
distribution ${\chi}_{Stern}^{2}$ and fixing the other unknown
parameters with the help of Stern data.

We now plot the graph for different confidence levels. Now our
best fit analysis with Stern observational data support the
theoretical range of the parameters. The 66\% (solid, blue), 90\%
(dashed, red) and 99\% (dashed, black) contours are plotted in
figures $1a$, $1b$, $1c$ and $1d$ for $\Omega_{m0}=0.28$,
$\Omega_{de0}=0.72$, $n=3.155$, $m=2$, $\omega=-25$, $f_0=0.9$,
$\phi_0=0.1$. The best fit values of ($\omega_{1},\omega_{0}$) are
tabulated in Table 2. Here the values of $\Omega_{m0}$ and
$\Omega_{de0}$ are motivated from the recent cosmological
observational data. The value of $\omega$ is chosen in accordance
with reference \cite{Kobayashi1}. The rest of the parameters
namely, $f_{0}$, $\phi_{0}$, $m$ and $n$ arises due to our
particular choice function of the scalar field. Due to their
secondary importance we consider them as free parameters and their
values are fine tuned in such a way so that effective closed
contours are obtained.

\subsection{Joint analysis with Stern+BAO datasets}
The method of joint analysis, the Baryon Acoustic Oscillation
(BAO) peak parameter value has been proposed by \cite{Eisenstein}
and we shall use their approach. Sloan Digital Sky Survey (SDSS)
survey  is one of the first redshift survey by which the BAO
signal has been directly detected at a scale $\sim$ 100 MPc. The
said analysis is actually the combination of angular diameter
distance and Hubble parameter at that redshift. This analysis is
independent of the measurement of $H_{0}$ and not containing any
particular dark energy. Here we examine the parameters
$\omega_{1}$ and $\omega_{0}$ for parametrized DE model from the
measurements of the BAO peak for low redshift (with range
$0<z<0.35$) using standard $\chi^{2}$ analysis. The error is
corresponding to the standard deviation, where we consider
Gaussian distribution. Low-redshift distance measurements is
lightly dependent on different cosmological parameters, the
equation of state of dark energy and have the ability to measure
the Hubble constant $H_{0}$ directly. The BAO peak parameter may
be defined by

\begin{equation}
{\cal
A}=\frac{\sqrt{\Omega_{m}}}{E(z_{1})^{1/3}}\left(\frac{1}{z_{1}}~\int_{0}^{z_{1}}
\frac{dz}{E(z)}\right)^{2/3}
\end{equation}
Here $E(z)=H(z)/H_{0}$ is the normalized Hubble parameter, the
redshift $z_{1}=0.35$ is the typical redshift of the SDSS sample
and the integration term is the dimensionless comoving distance to
the to the redshift $z_{1}$ The value of the parameter ${\cal A}$
for the flat model of the universe is given by ${\cal A}=0.469\pm
0.017$ using SDSS data \cite{Eisenstein} from luminous red
galaxies survey. Now the $\chi^{2}$ function for the BAO
measurement can be written as

\begin{equation}
\chi^{2}_{BAO}=\frac{({\cal A}-0.469)^{2}}{(0.017)^{2}}
\end{equation}
Now the total joint data analysis (Stern+BAO) for the $\chi^{2}$
function may be defined by

\begin{equation}
\chi^{2}_{total}=\chi^{2}_{Stern}+\chi^{2}_{BAO}
\end{equation}
According to our analysis the joint scheme(Stern+BAO) gives the
best fit values of ($\omega_{1},\omega_{0}$) in Table 3. Finally
we draw the contours for the 66\% (solid,blue), 90\% (dashed, red)
and 99\%(dashed, black) confidence limits depicted in figures
$2a$, $2b$, $2c$ and $2d$ for $\Omega_{m0}=0.28$,
$\Omega_{de0}=0.72$, $\alpha=0.001$, $n=0.5$,$m=10$, $\omega=-3$,
$w_m=0.03$,
$f_0=0.01$, $\phi_0=0.01$.\\

\subsection{Joint analysis with Stern+BAO+CMB datasets}
One interesting geometrical probe of dark energy can be determined
by the angular scale of the first acoustic peak through angular
scale of the sound horizon at the surface of last scattering which
is encoded in the CMB power spectrum Cosmic Microwave Background
(CMB) shift parameter is defined by \cite{Bond1, Efstathiou1,
Nessaeris1}. It is not sensitive with respect to perturbations but
are suitable to constrain model parameter. The CMB power spectrum
first peak is the shift parameter which is given by

\begin{equation}
{\cal R}=\sqrt{\Omega_{m}} \int_{0}^{z_{2}} \frac{dz}{E(z)}
\end{equation}
where $z_{2}$ is the value of redshift at the last scattering
surface. From WMAP7 data of the work of Komatsu et al.
\cite{Komatsu1} the value of the parameter has obtained as ${\cal
R}=1.726\pm 0.018$ at the redshift $z=1091.3$. Now the $\chi^{2}$
function for the CMB measurement can be written as

\begin{equation}
\chi^{2}_{CMB}=\frac{({\cal R}-1.726)^{2}}{(0.018)^{2}}
\end{equation}
Now when we consider three cosmological tests together, the total
joint data analysis (Stern+BAO+CMB) for the $\chi^{2}$ function
may be defined by

\begin{equation}
\chi^{2}_{TOTAL}=\chi^{2}_{Stern}+\chi^{2}_{BAO}+\chi^{2}_{CMB}
\end{equation}
Now the best fit values of ($\omega_{1},\omega_{0}$) for joint
analysis of BAO and CMB with Stern observational data support the
theoretical range of the parameters given in Table 4. The 66\%
(solid, blue), 90\% (dashed, red) and 99\% (dashed, black)
contours are plotted in figures $3a$, $3b$, $3c$ and $3d$ for
$\Omega_{m0}=0.28$, $\Omega_{de0}=0.72$, $\alpha=0.001$,
$n=0.5$,$m=10$, $\omega=-3$, $w_m=0.03$, $f_0=0.01$,
$\phi_0=0.01$.
\vspace{1mm}
\begin{figure}[h]
\includegraphics[height=1.6in]{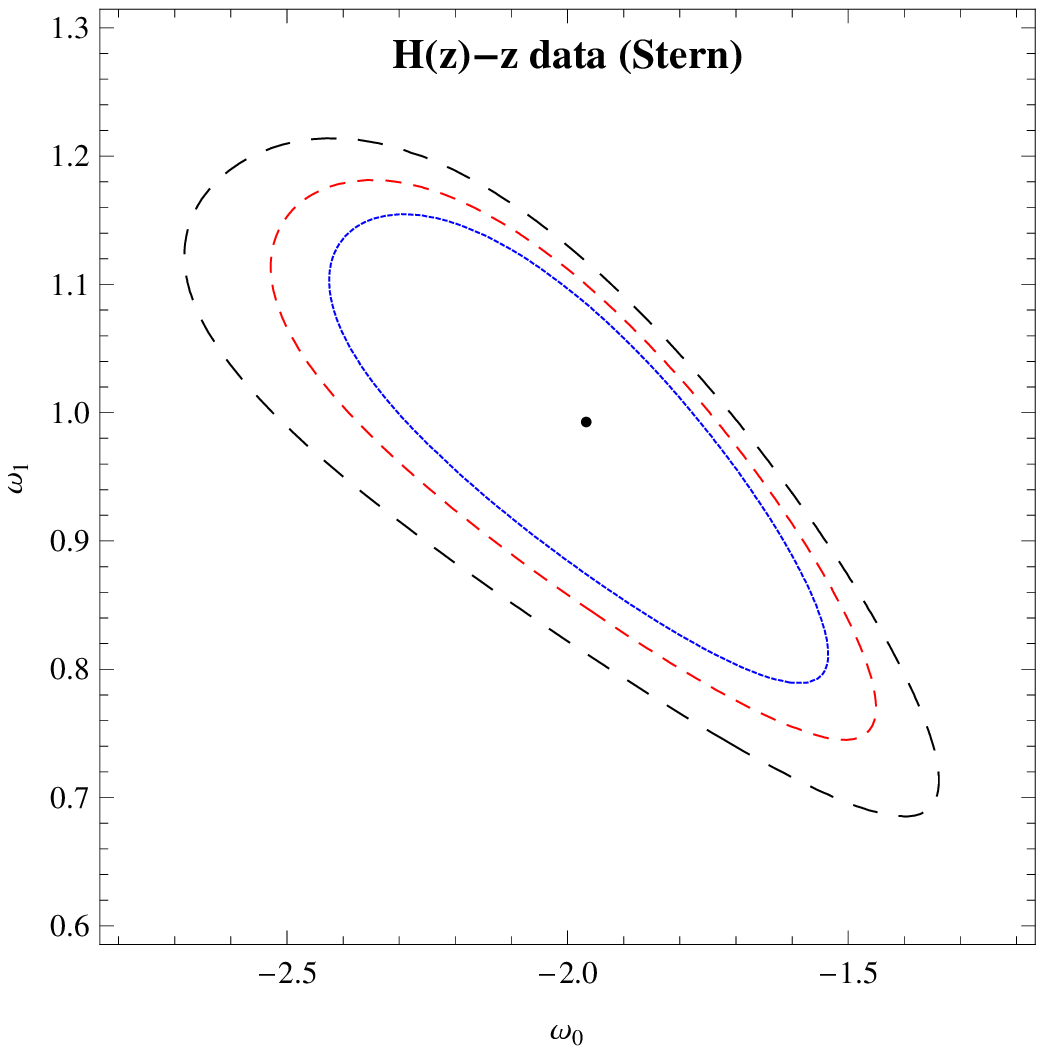}~~~~\includegraphics[height=1.6in]{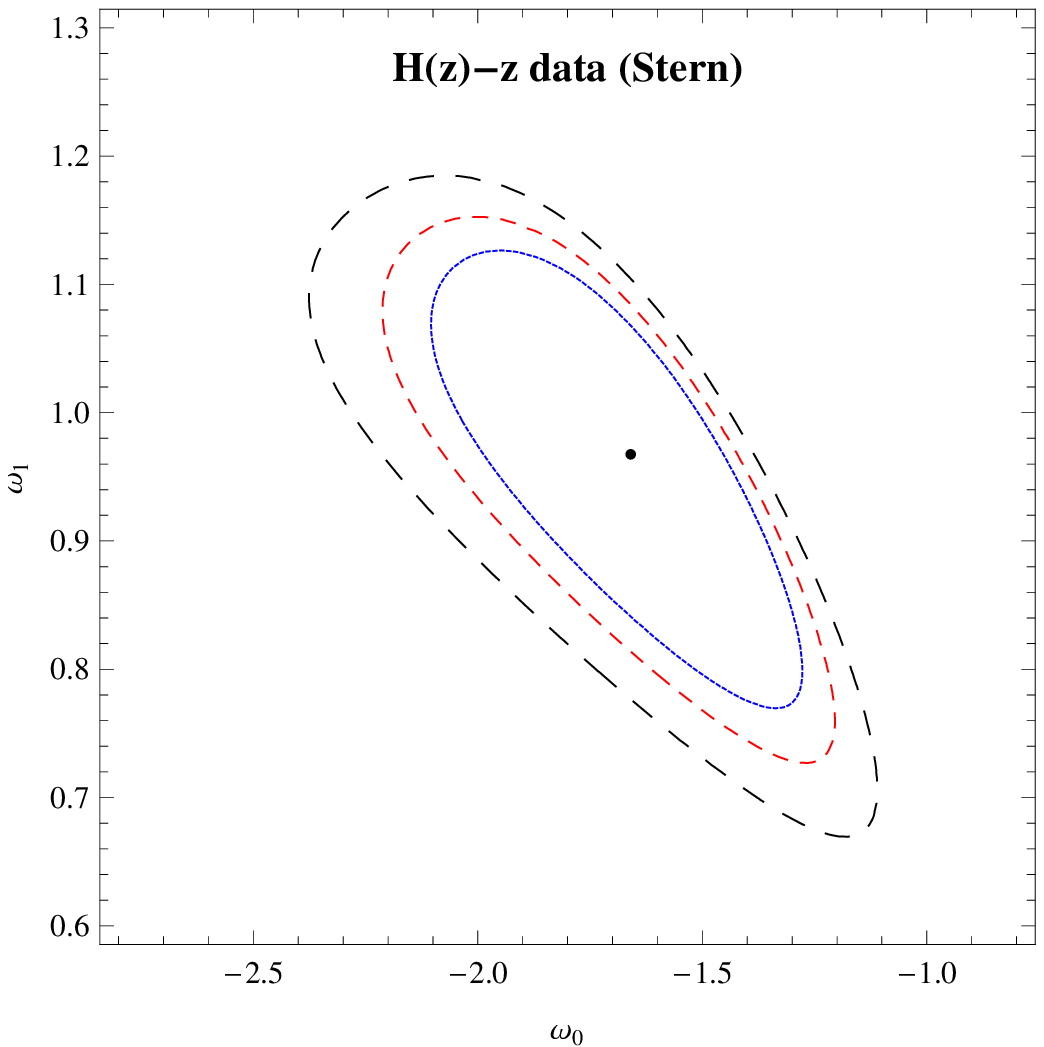}~~~~
\includegraphics[height=1.6in]{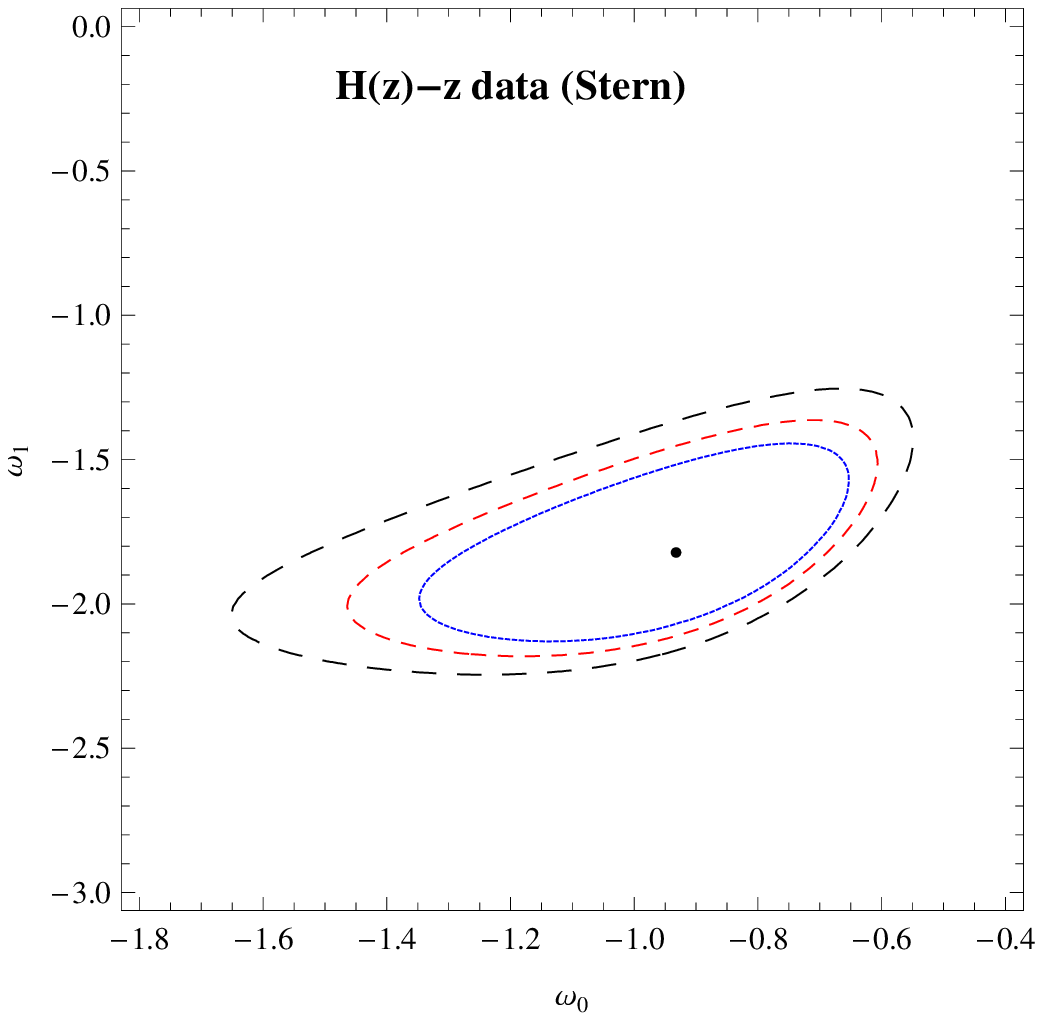}~~~~\includegraphics[height=1.6in]{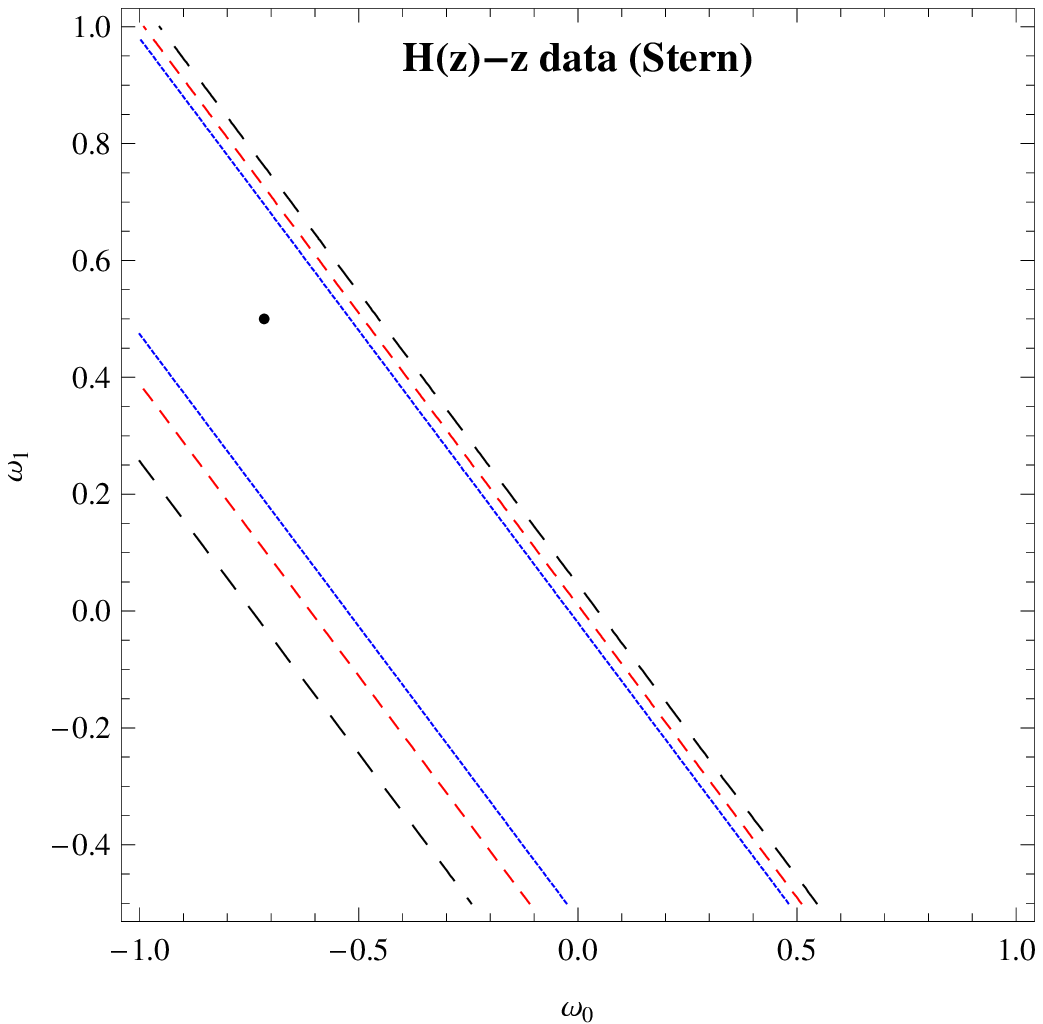}~~~~\\
\vspace{1mm}
~~~~~~~~~~~~~Fig.1a~~~~~~~~~~~~~~~~~~~~~~~~~~~~~~~~~~~~~~~~~~Fig.1b~~~~~~~~~~~~~~~~~~~~~~~~~~~
~~~~~~~~~~~~~Fig.1c~~~~~~~~~~~~~~~~~~~~~~~~~~~~~~~~~~~~~~~~~~Fig.1d~~~~~~~~~~~~~\\
\vspace{2mm} \textit{\textbf{Figs 1a, 1b, 1c and 1d} show the
variation of $\omega_{0}$ with $\omega_{1}$ for different
confidence levels for linear, CPL, JBP and logarithmic models
respectively. The $66\%$ (solid, blue, the innermost contour),
$90\%$ (dashed, red, next to the innermost contour), and $99\%$
(dashed, black, the outermost contour) contours are plotted in
these figures for the $H(z)-z$ (Stern) analysis.}
\end{figure}
\vspace{1mm}
\begin{figure}[h]
\includegraphics[height=1.6in]{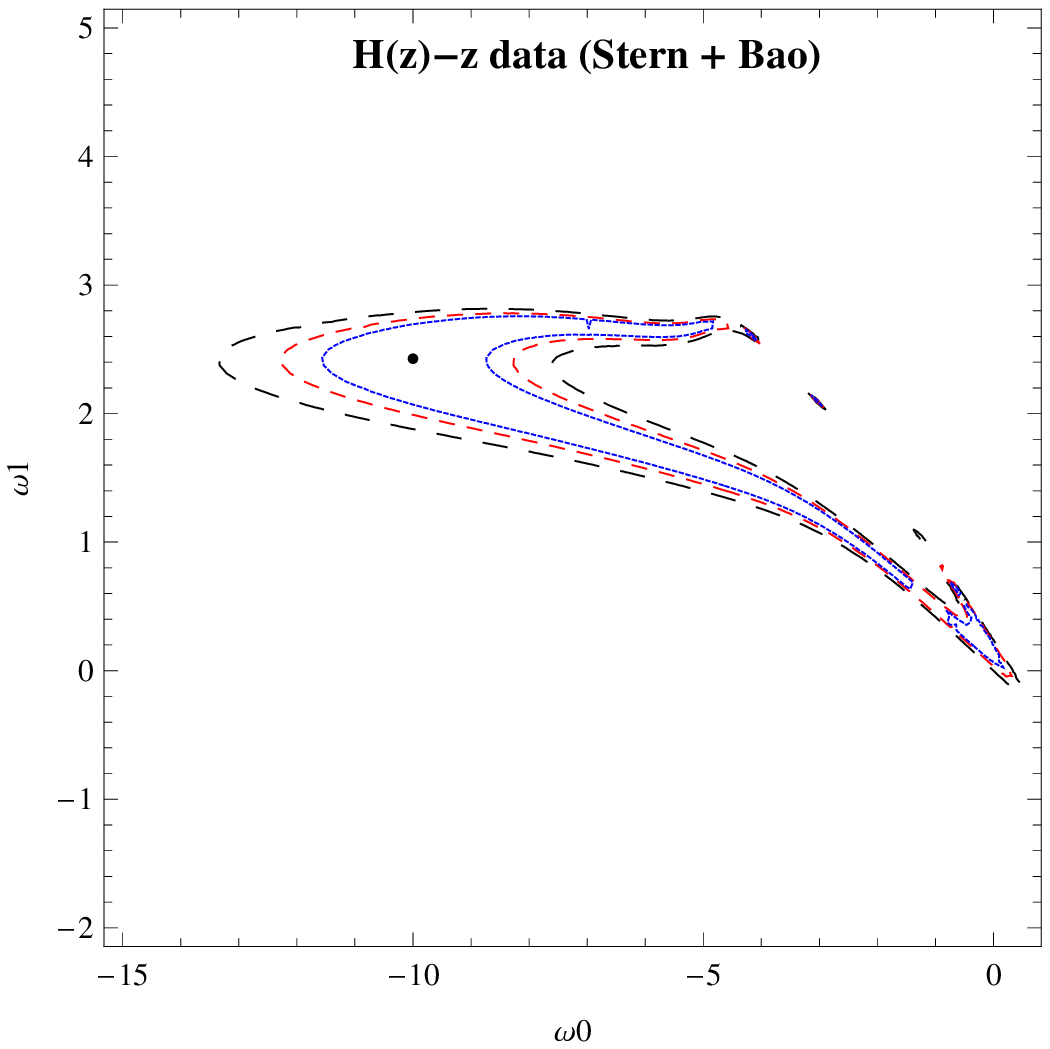}~~~~\includegraphics[height=1.6in]{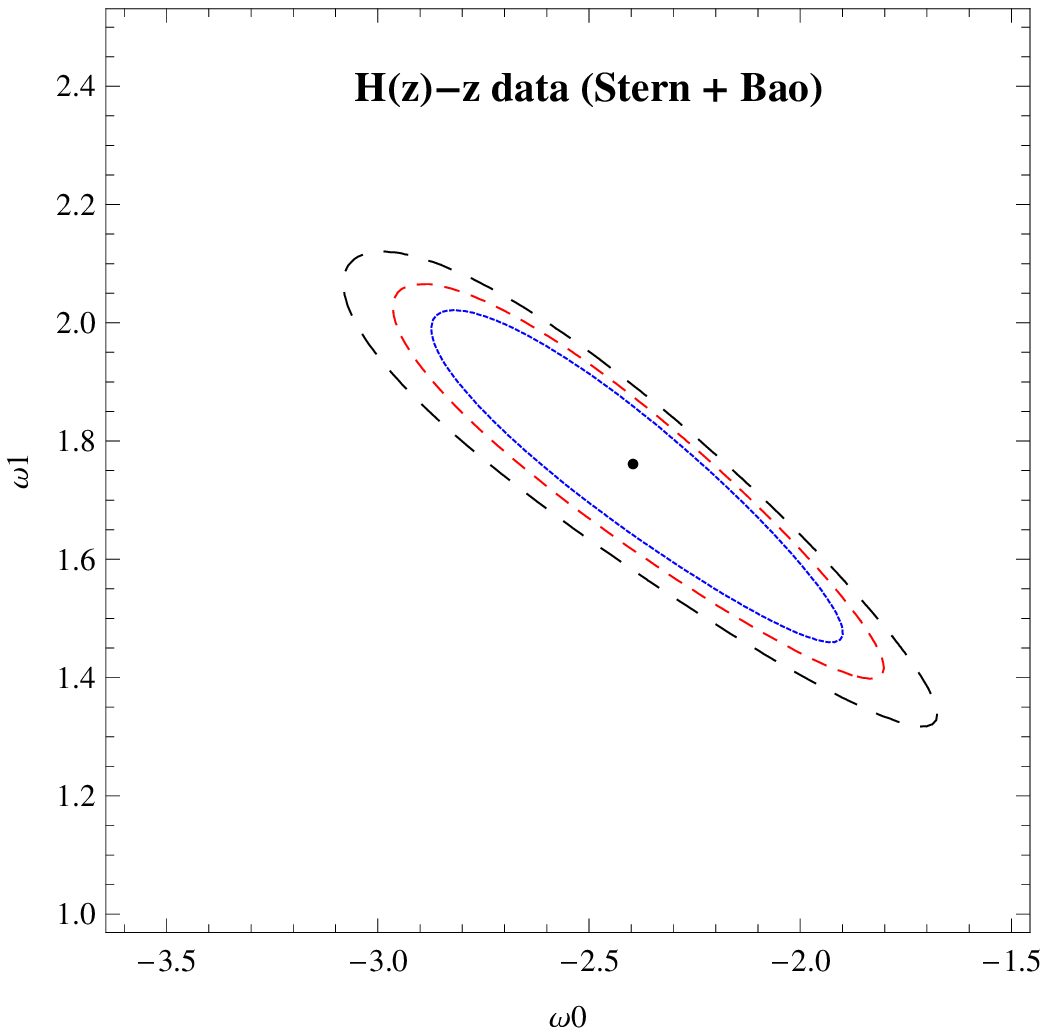}~~~~
\includegraphics[height=1.6in]{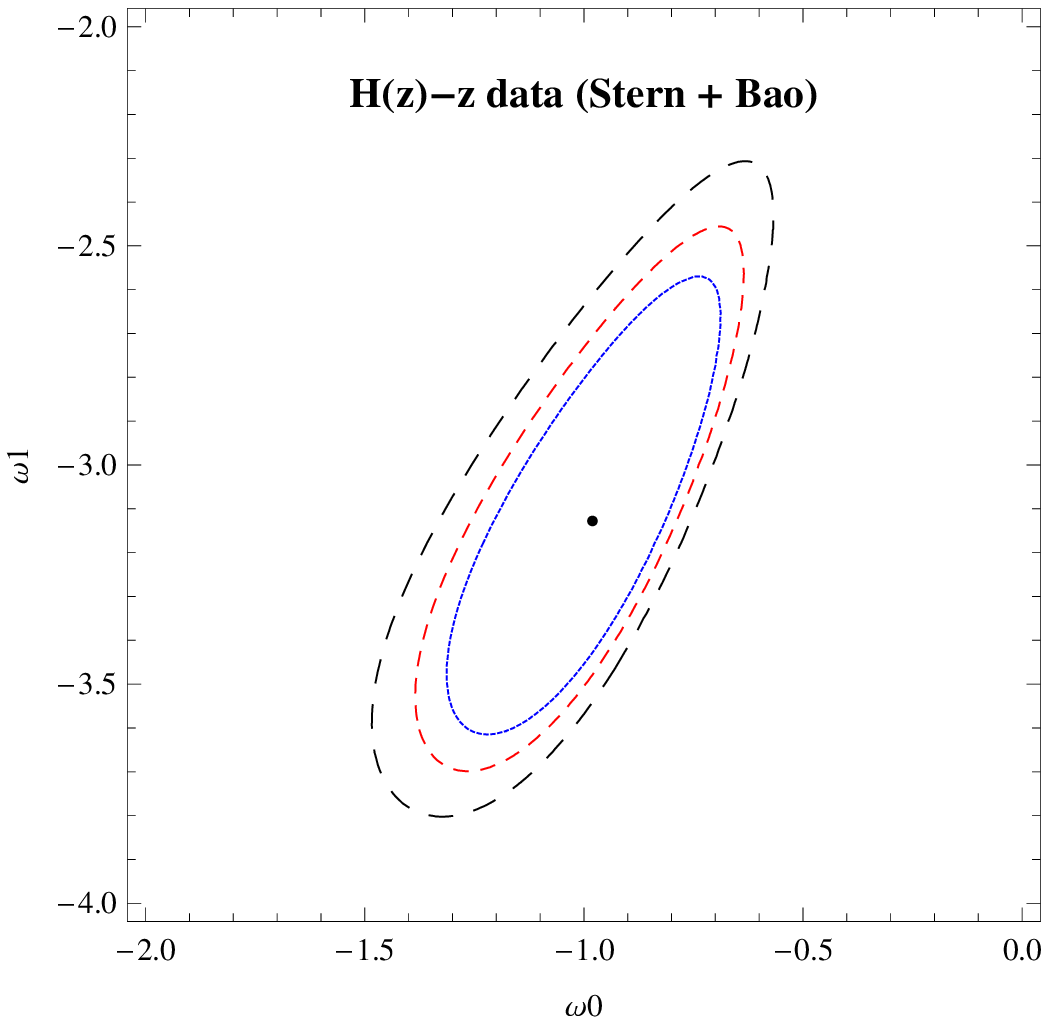}~~~~\includegraphics[height=1.6in]{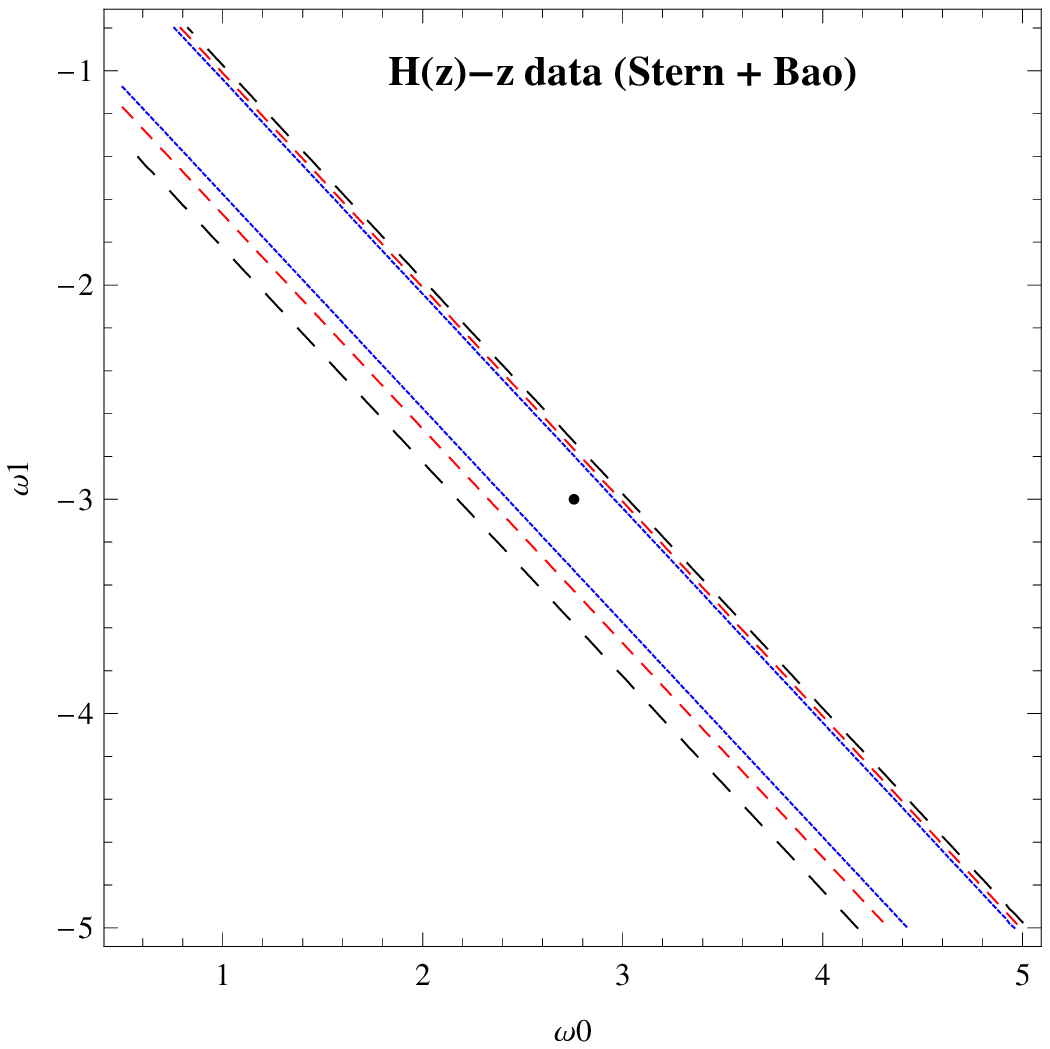}~~~~\\
\vspace{1mm}
~~~~~~~~~~~~~Fig.2a~~~~~~~~~~~~~~~~~~~~~~~~~~~~~~~~~~~~~~~~~~Fig.2b~~~~~~~~~~~~~~~~~~~~~~~~~~~
~~~~~~~~~~~~~Fig.2c~~~~~~~~~~~~~~~~~~~~~~~~~~~~~~~~~~~~~~~~~~Fig.2d~~~~~~~~~~~~~\\
\vspace{2mm} \textit{\textbf{Figs 2a, 2b, 2c and 2d} show the
variation of $\omega_{0}$ with $\omega_{1}$ for different
confidence levels for linear, CPL, JBP and logarithmic models
respectively. The $66\%$ (solid, blue, the innermost contour),
$90\%$ (dashed, red, next to the innermost contour), and $99\%$
(dashed, black, the outermost contour) contours are plotted in
these figures for the $H(z)–z$ (Stern+BAO) joint analysis.}
\end{figure}
\vspace{1mm}
\begin{figure}[h]
\includegraphics[height=1.6in]{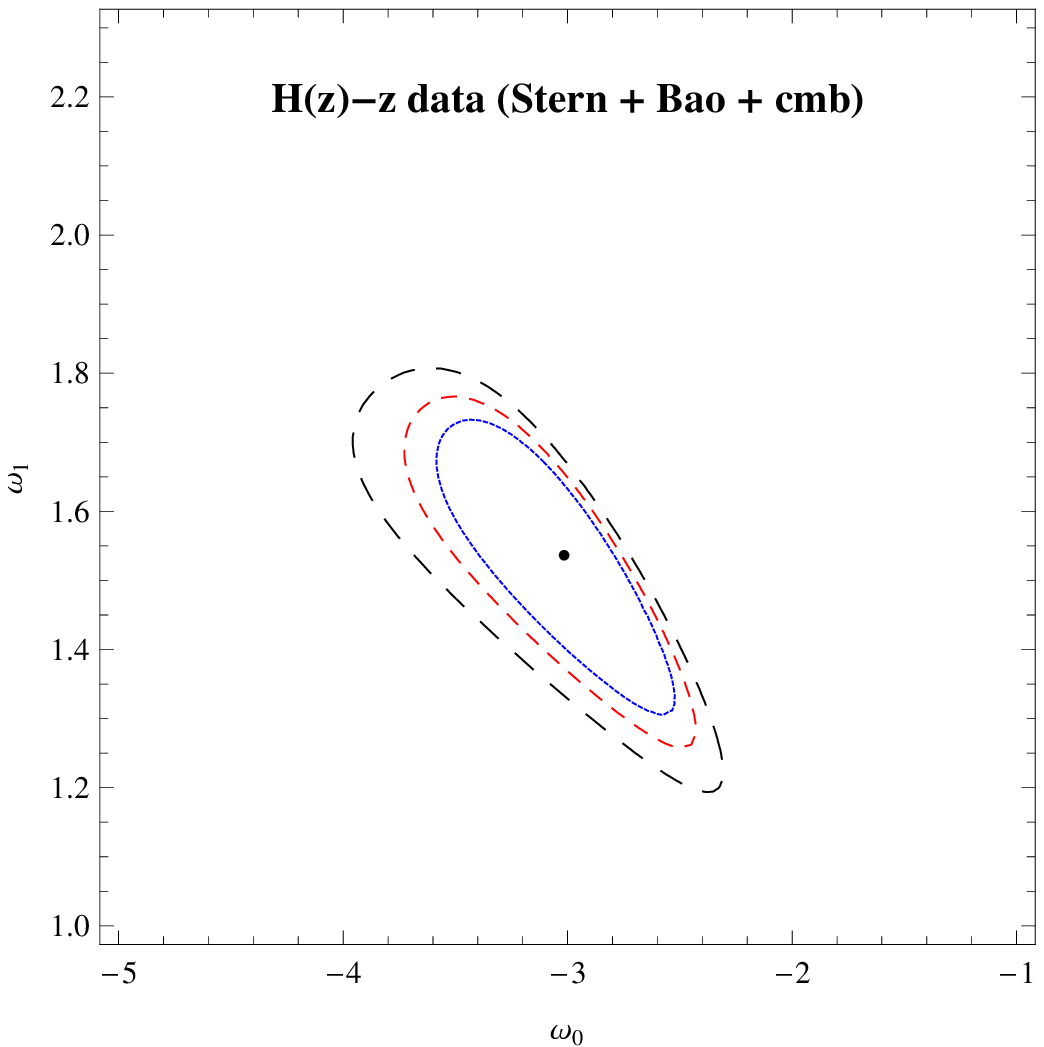}~~~~\includegraphics[height=1.6in]{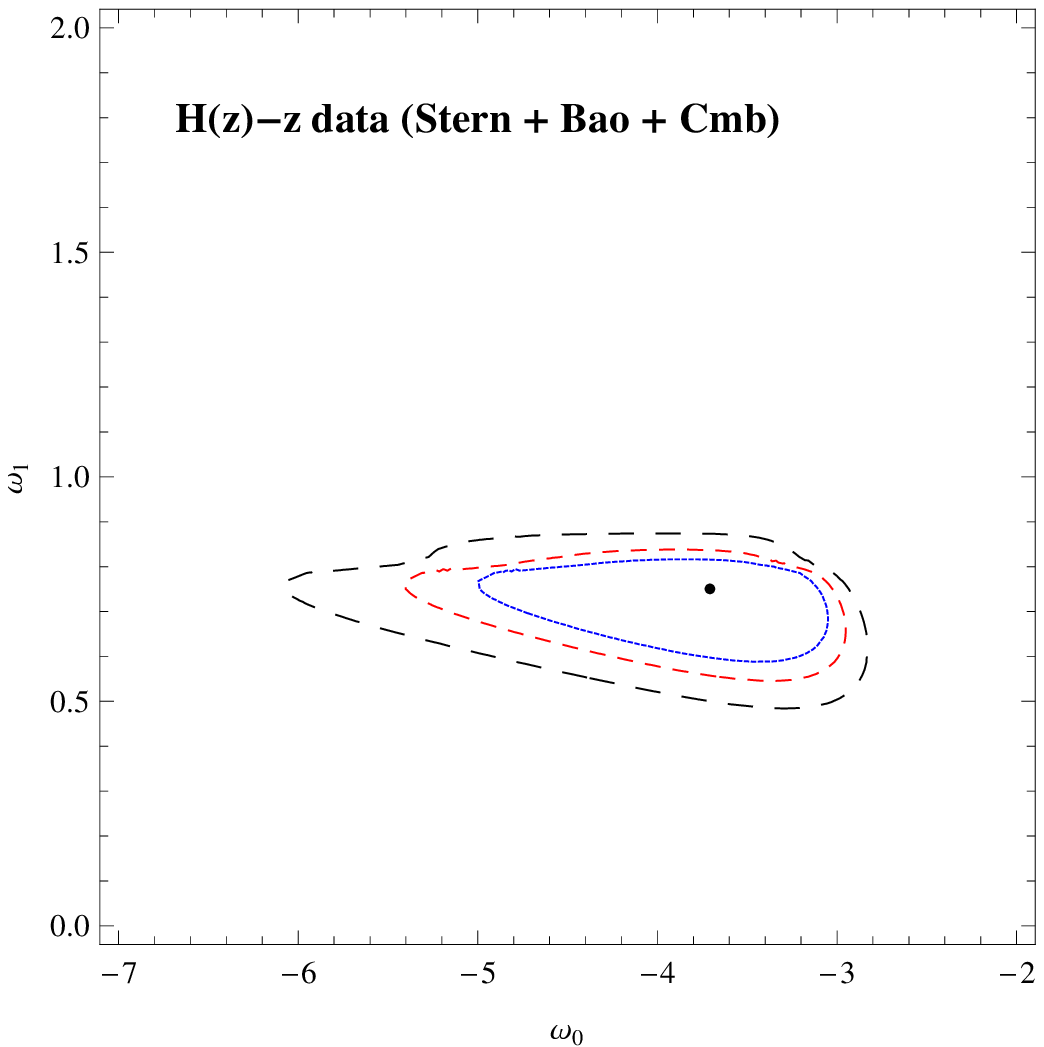}~~~~
\includegraphics[height=1.6in]{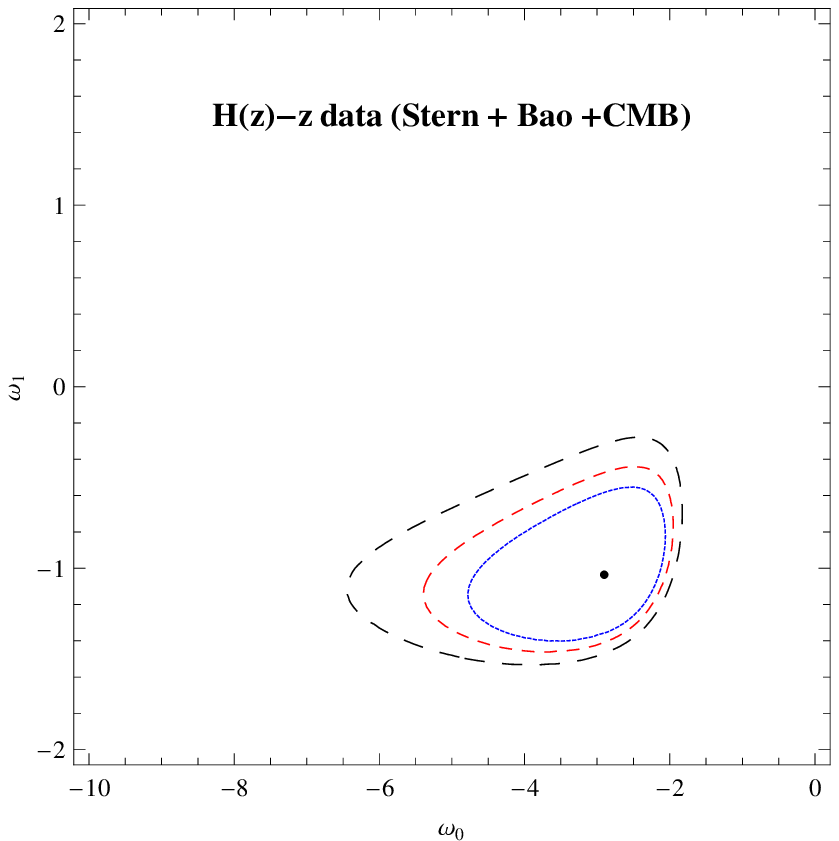}~~~~\includegraphics[height=1.6in]{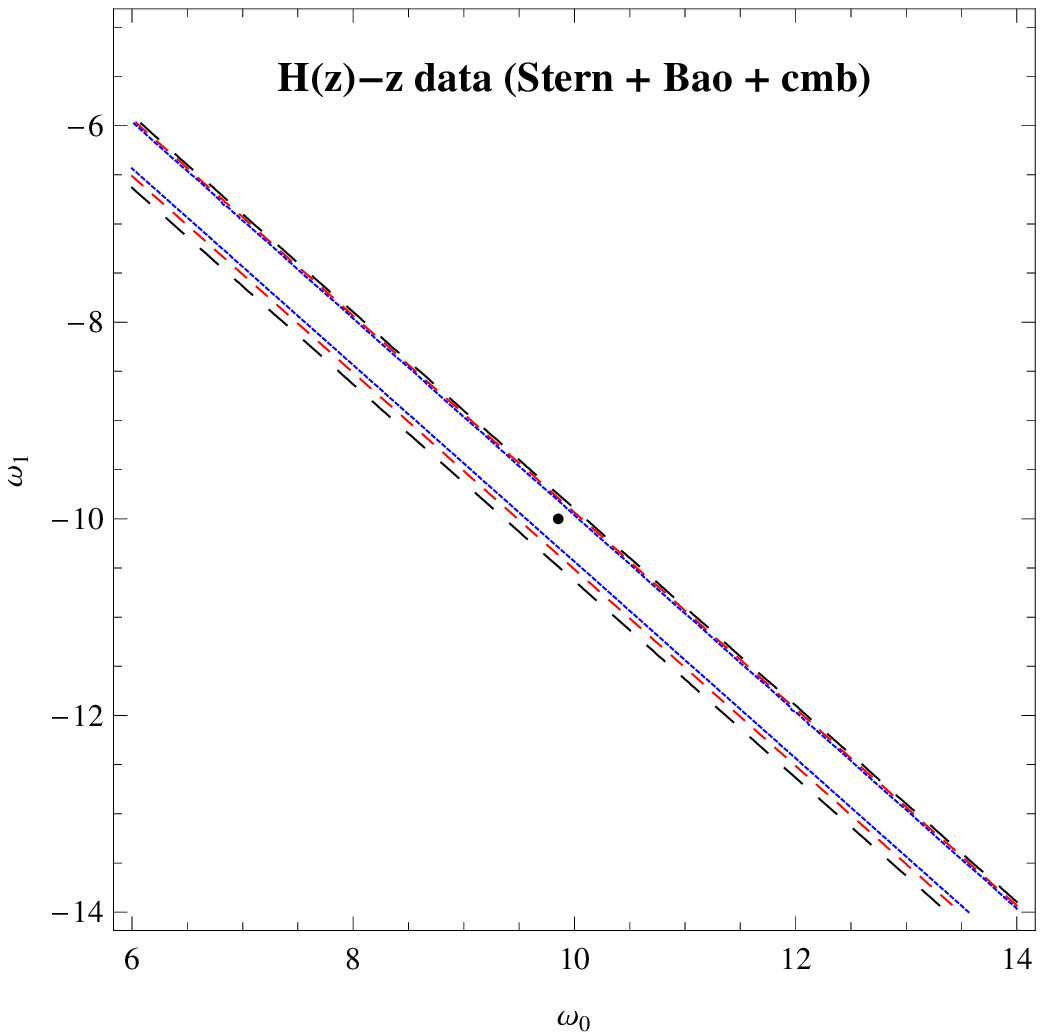}~~~~\\
\vspace{1mm}
~~~~~~~~~~~~~Fig.3a~~~~~~~~~~~~~~~~~~~~~~~~~~~~~~~~~~~~~~~~~~Fig.3b~~~~~~~~~~~~~~~~~~~
~~~~~~~~~~~~~Fig.3c~~~~~~~~~~~~~~~~~~~~~~~~~~~~~~~~~~~~~~~~~~Fig.3d~~~~~~~~~~~~~\\
\vspace{2mm} \textit{\textbf{Figs 3a, 3b, 3c and 3d} show the
variation of $\omega_{0}$ with $\omega_{1}$ for different
confidence levels for linear, CPL, JBP and logarithmic models
respectively. The $66\%$ (solid, blue, the innermost contour),
$90\%$ (dashed, red, next to the innermost contour), and $99\%$
(dashed, black, the outermost contour) contours are plotted in
these figures for the $H(z)–z$ (Stern+BAO+CMB) joint analysis.}
\end{figure}
\vspace{1mm}
\begin{figure}[h]
\includegraphics[height=1.6in]{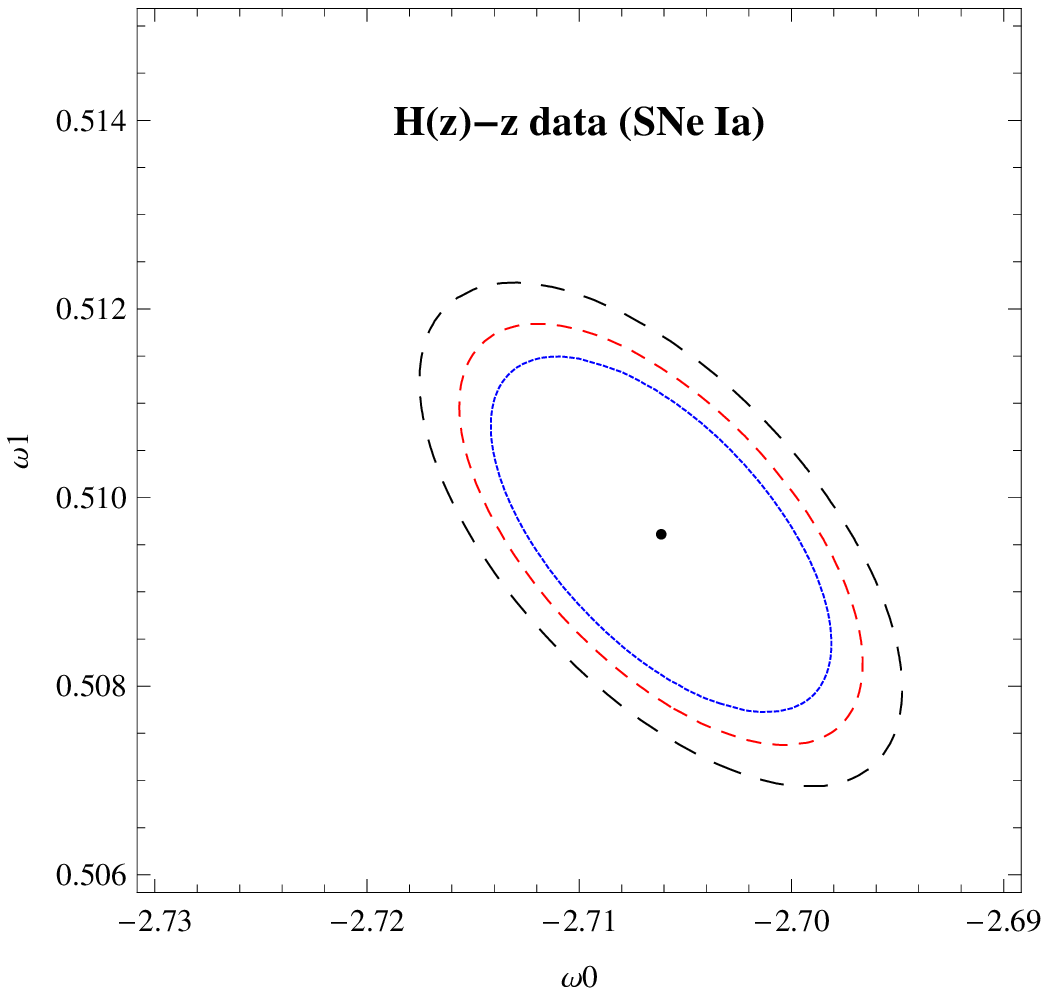}~~~\includegraphics[height=1.6in]{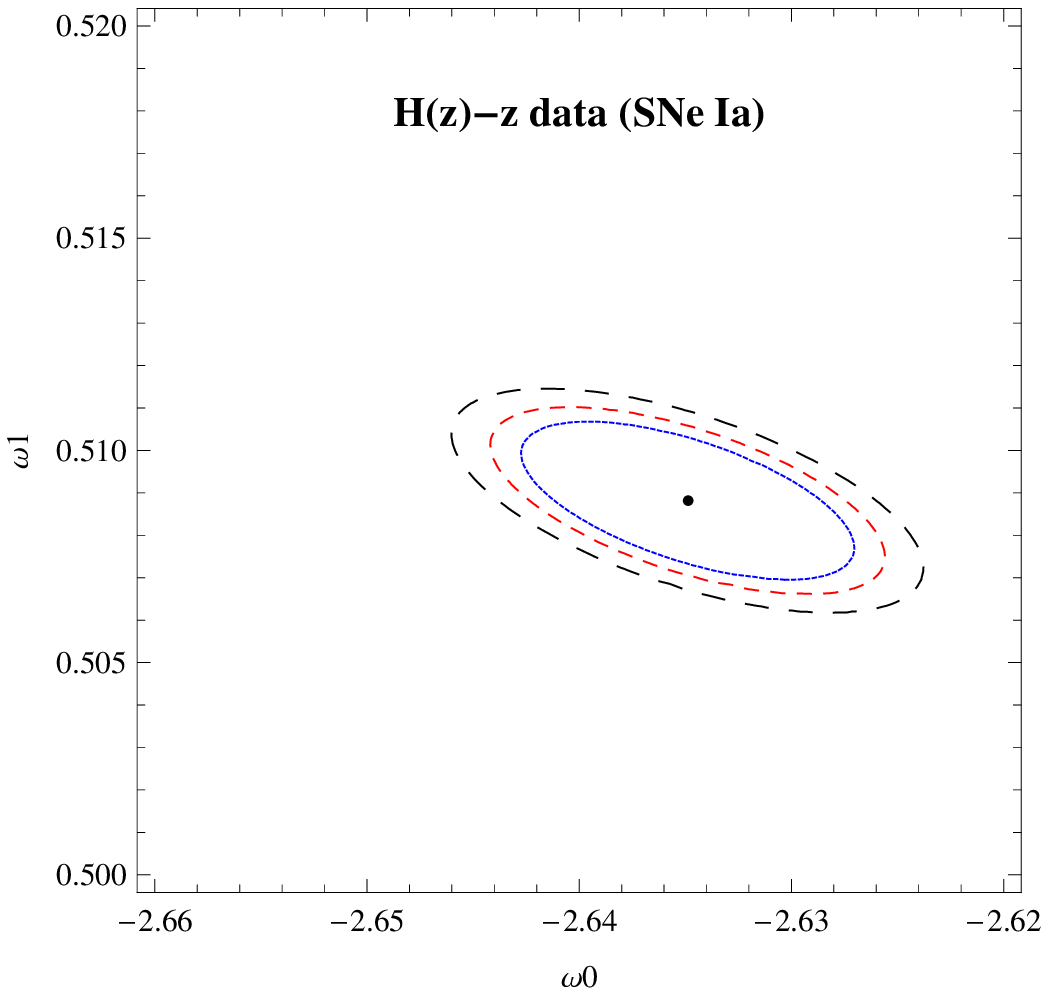}~~
\includegraphics[height=1.6in]{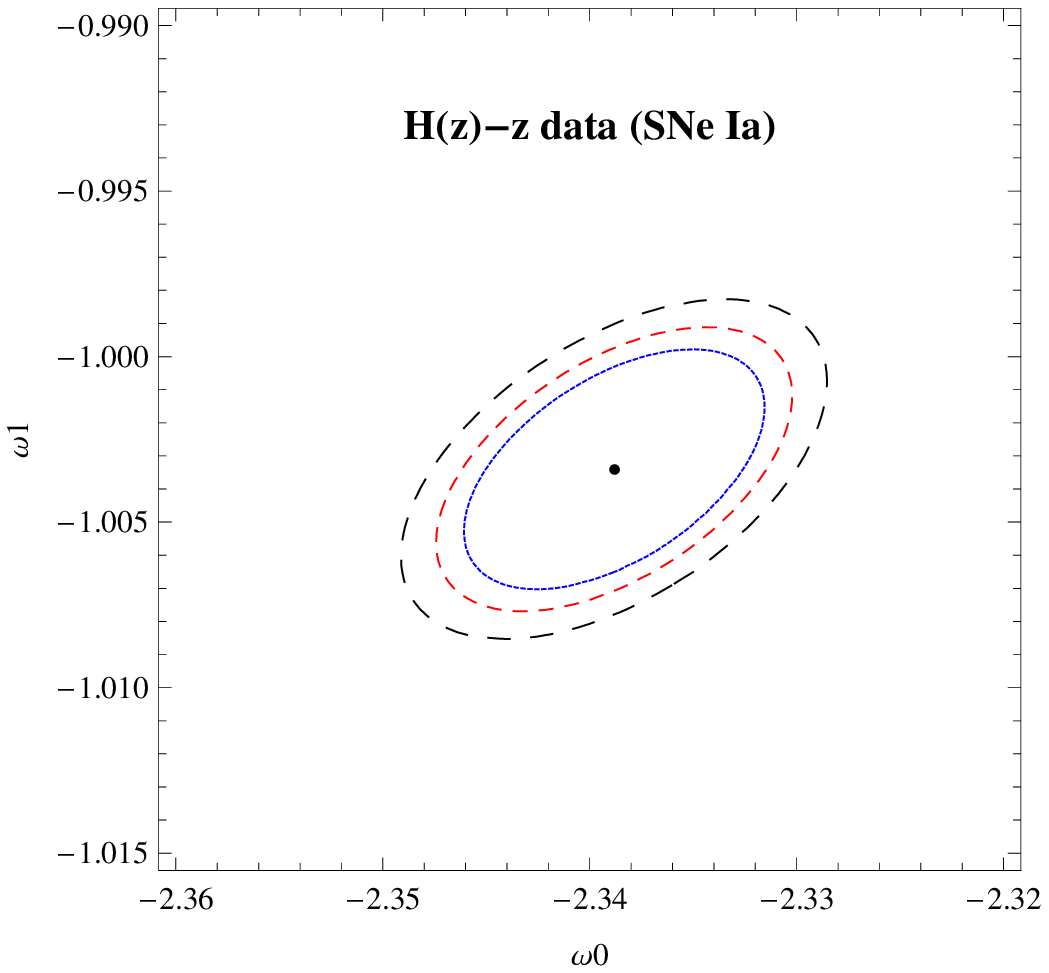}~~~\includegraphics[height=1.6in]{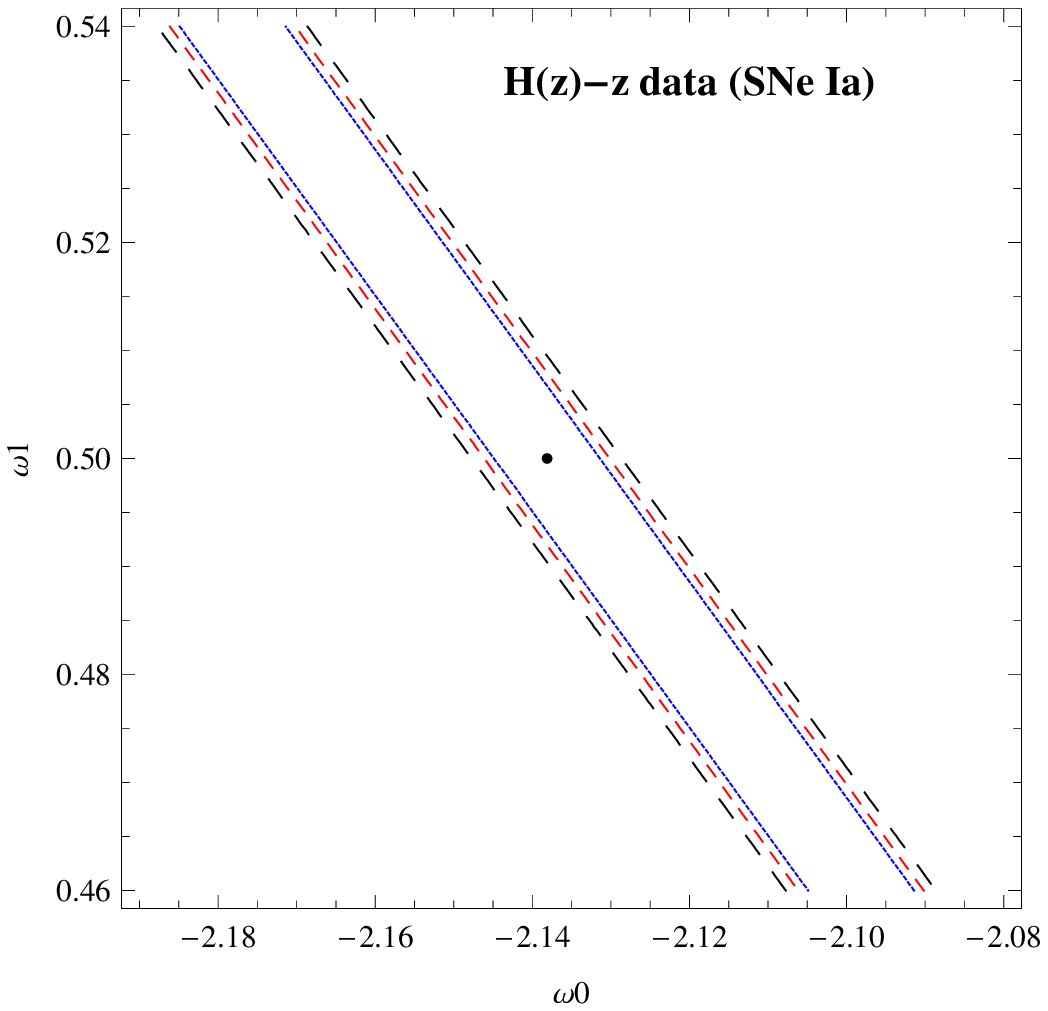}~~~\\
\vspace{1mm}
~~~~~~~~~~~~~Fig.4a~~~~~~~~~~~~~~~~~~~~~~~~~~~~~~~~~~~~~~~~~~Fig.4b~~~~~~~~~~~~~~~~~~~~~~~~~~~
~~~~~~~~~~~~~Fig.4c~~~~~~~~~~~~~~~~~~~~~~~~~~~~~~~~~~~~~~~~~~Fig.4d~~~~~~~~~~~~~\\
\vspace{2mm} \textit{\textbf{Figs 4a, 4b, 4c and 4d} show the
variation of $\omega_{0}$ with $\omega_{1}$ for different
confidence levels for linear, CPL, JBP and logarithmic models
respectively. The $66\%$ (solid, blue, the innermost contour),
$90\%$ (dashed, red, next to the innermost contour), and $99\%$
(dashed, black, the outermost contour) contours are plotted in
these figures for the $H(z)–z$ of SNe Type Ia 292 data.}
\end{figure}
\vspace{1mm}
\begin{figure}[h]
\includegraphics[height=2in]{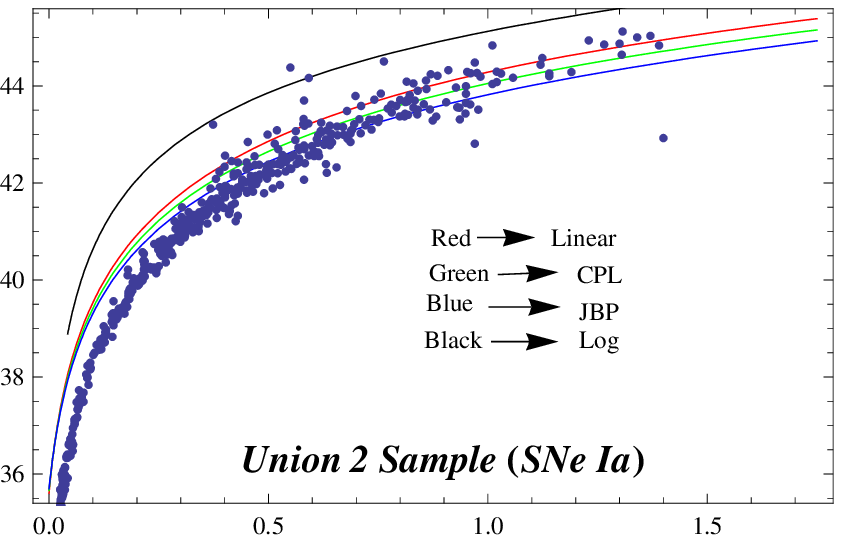}~~~~~~~~~\includegraphics[height=2in]{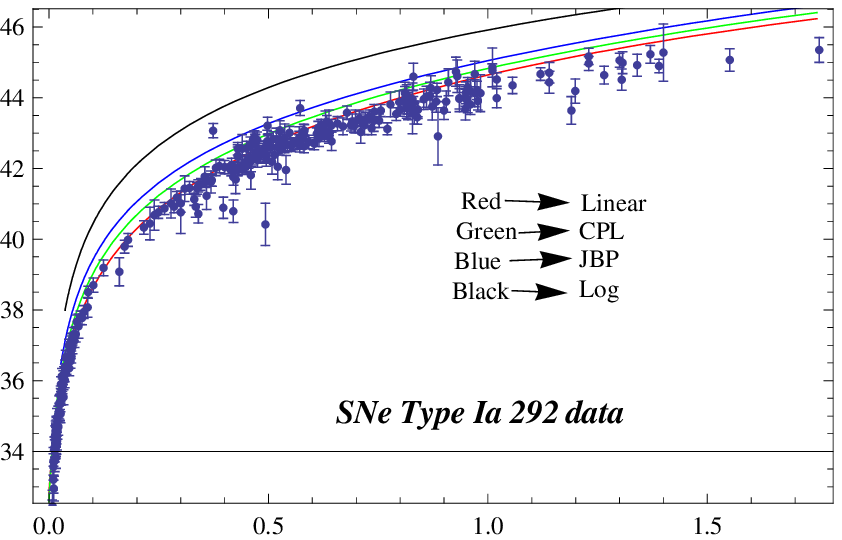}~~~~\\
\vspace{1mm}
~~~~~~~~~~~~~~~~~~Fig.5a~~~~~~~~~~~~~~~~~~~~~~~~~~~~~~~~~~~~~~~~~~~~~~~~~~~~~~~~~~~~~~Fig.5b~~~~~~~~~~~~~\\
\vspace{2mm} \textit{\textbf{Figs 5a and 5b} show the variation of
$\mu(z)$ vs $z$ for linear, CPL, JBP and logarithmic
parametrizations (solid lines) and their comparison with
observational data (blue dots). The dots denote the Union2 sample
data in fig 5a and the SNe Type Ia 292 data in fig 5b.}
\end{figure}

\vspace{2mm}
\[
\begin{tabular}{|c|c|c|c|}
\hline
  ~~~~~~\textit{DE models}~~~~~~~~~~~~~~&~~~~~~~~~\textit{Best fit values}~~~~~~&~~~~~~~~~~~~~~~~~~~~~~~~~&~~~~~~~~~~~~~~~~~~~~~~~~~\\
  \hline
  ~~~~~~~~~~~~~~~~~~~~~~~~~~~~~~~~~~~~~~&~~~~~~~$\omega_{0}$ ~~~~~~~~&~~~~~~~~$\omega_{1}$~~~~~&~~~~~$\chi^{2}_{min}$~~~~~~\\
  \hline
  ~~~~~~$Linear$ ~~~~~&~~~~~$-1.96664$ ~~~~~~&~~~$0.992802$~~~~~&~~~~~~~~$11.4933$~~~~~~~~~~~\\
  \hline
  ~~~~~~$CPL$ ~~~~~&~~~~~$-1.65925$ ~~~~~~&~~~$0.967584$~~~~~&~~~~~~~~$8.4682$~~~~~~~~~~~\\
  \hline
   ~~~~~~$JBP$ ~~~~~&~~~~~$-0.932444$ ~~~~~~&~~~$-1.82202$~~~~~&~~~~~~~~$7.88769$~~~~~~~~~~~\\
   \hline
   ~~~~~~$Log$ ~~~~~&~~~~~$-0.716018$ ~~~~~~&~~~$0.5$~~~~~&~~~~~~~~$127.572$~~~~~~~~~~~\\
   \hline
\end{tabular}
\]
{\bf Table 2:} STERN: The best fit values of $\omega_{0}$ and
$\omega_{1}$ and the minimum values of $\chi^{2}$ for different
redshift parametrization models. \vspace{1mm}
\[
\begin{tabular}{|c|c|c|c|}
\hline
  ~~~~~~\textit{DE models}~~~~~~~~~~~~~~&~~~~~~~~~\textit{Best fit values}~~~~~~&~~~~~~~~~~~~~~~~~~~~~~~~~&~~~~~~~~~~~~~~~~~~~~~~~~~\\
  \hline
  ~~~~~~~~~~~~~~~~~~~~~~~~~~~~~~~~~~~~~~&~~~~~~~$\omega_{0}$ ~~~~~~~~&~~~~~~~~$\omega_{1}$~~~~~&~~~~~$\chi^{2}_{min}$~~~~~~\\
  \hline
  ~~~~~$Linear$ ~~~&~~~~~$-10$ ~~~~~~&~~~$2.42671$~~~~~&~~~~~$928.664$~~~~~~\\
  \hline
  ~~~~~$CPL$ ~~~&~~~~~$-2.39571$ ~~~~~~&~~~$1.76133$~~~~~&~~~~~$773.265$~~~~~~\\
  \hline
  ~~~~~$JBP$ ~~~&~~~~~$-0.980504$ ~~~~~~&~~~$-3.12768$~~~~~&~~~~~$781.012$~~~~~~\\
  \hline
  ~~~~~$Log$ ~~~&~~~~~$2.75676$ ~~~~~~&~~~$-3$~~~~~&~~~~~$903.023$~~~~~~\\
  \hline
  \end{tabular}
\]
{\bf Table 3:} STERN+BAO: The best fit values of $\omega_{0}$ and
$\omega_{1}$ and the minimum values of $\chi^{2}$ for different
redshift parametrization models.
\[
\begin{tabular}{|c|c|c|c|}
\hline
  ~~~~~~\textit{DE models}~~~~~~~~~~~~~~&~~~~~~~~~\textit{Best fit values}~~~~~~&~~~~~~~~~~~~~~~~~~~~~~~~~&~~~~~~~~~~~~~~~~~~~~~~~~~\\
  \hline
  ~~~~~~~~~~~~~~~~~~~~~~~~~~~~~~~~~~~~~~&~~~~~~~$\omega_{0}$ ~~~~~~~~&~~~~~~~~$\omega_{1}$~~~~~&~~~~~$\chi^{2}_{min}$~~~~~~\\
  \hline
  ~~~~$Linear$~~&~~~~~~$-3.01522$ ~~~~~~&~~~~$1.53659$~~~~~&~~~~~$9969.52$~~~~~~\\
  \hline
  ~~~~$CPL$~~&~~~~~~$-3.70734$ ~~~~~~&~~~~$0.750595$~~~~~&~~~~~$11645.9$~~~~~~\\
  \hline
  ~~~~$JBP$~~&~~~~~~$-2.9022$ ~~~~~~&~~~~$-1.03615$~~~~~&~~~~~$10400.9$~~~~~~\\
  \hline
  ~~~~$Log$~~&~~~~~~$9.85277$ ~~~~~~&~~~~$-10$~~~~~&~~~~~$10105.5$~~~~~~\\
   \hline
\end{tabular}
\]
{\bf Table 4:} STERN+BAO+CMB: The best fit values of $\omega_{0}$
and $\omega_{1}$ and the minimum values of $\chi^{2}$ for
different redshift parametrization models.\vspace{1mm}
\[
\begin{tabular}{|c|c|c|c|}
\hline
  ~~~~~~\textit{DE models}~~~~~~~~~~~~~~&~~~~~~~~~\textit{Best fit values}~~~~~~&~~~~~~~~~~~~~~~~~~~~~~~~~&~~~~~~~~~~~~~~~~~~~~~~~~~\\
  \hline
  ~~~~~~~~~~~~~~~~~~~~~~~~~~~~~~~~~~~~~~&~~~~~~~$\omega_{0}$ ~~~~~~~~&~~~~~~~~$\omega_{1}$~~~~~&~~~~~$\chi^{2}_{min}$~~~~~~\\
  \hline
  ~~~~$Linear$~~&~~~~~~$-2.70613$ ~~~~~~&~~~~$0.509612$~~~~~&~~~~~$879522.25$~~~~~~\\
  \hline
  ~~~~$CPL$~~&~~~~~~$-2.63487$ ~~~~~~&~~~~$0.508819$~~~~~&~~~~~$867545.51$~~~~~~\\
  \hline
  ~~~~$JBP$~~&~~~~~~$-2.3388$ ~~~~~~&~~~~$-1.00341$~~~~~&~~~~~$847931.41$~~~~~~\\
  \hline
  ~~~~$Log$~~&~~~~~~$-2.13811$ ~~~~~~&~~~~$0.5$~~~~~&~~~~~$1.32293 \times 10^6$~~~~~~\\
   \hline
\end{tabular}
\]
{\bf Table 5:} H(z)-z SNe Type Ia 292: The best fit values of
$\omega_{0}$ and $\omega_{1}$ and the minimum values of $\chi^{2}$
for different redshift parametrization models.

\section{Redshift-magnitude observations from supernovae type Ia Union2 sample (from Amanullah et al. 2010)}
The Supernova Type Ia experiments provided the main evidence for
the existence of dark energy. Since 1995, two teams of High-$z$
Supernova Search and the Supernova Cosmology Project have
discovered several type Ia supernovas at the high redshifts
\cite{Perlmutter, Riess, Riess1, Perlmutter1}. The observations
directly measure the distance modulus of a Supernovae and its
redshift $z$ \cite{Riess2,Kowalaski1}. Now, take recent
observational data, including SNe Ia which consists of 557 data
points and belongs to the Union2 sample \cite{Amanullah1}. From
the observations, the luminosity distance $d_{L}(z)$ determines
the dark energy density and is defined by

\begin{equation}
d_{L}(z)=(1+z)H_{0}\int_{0}^{z}\frac{dz'}{H(z')}
\end{equation}
and the distance modulus (distance between absolute and apparent
luminosity of a distance object) for Supernovas is given by

\begin{equation}
\mu(z)=5\log_{10} \left[\frac{d_{L}(z)/H_{0}}{1~MPc}\right]+25
\end{equation}
The best fit of distance modulus as a function $\mu(z)$ of
redshift $z$ for our theoretical models and the Supernova Type Ia
Union2 sample are drawn in figure 5a for our best fit values of
$\omega_{1}$ and $\omega_{0}$ for (Stern) + BAO + CMB joint
analysis as $\Omega_{m0}=0.28$, $\Omega_{de0}=0.72$,
$\alpha=0.001$, $n=0.5$,$m=10$, $\omega=-3$, $w_m=0.03$,
$f_0=0.01$, $\phi_0=0.01$. Similarly the best fit trajectories for
our theoretical models are obtained and compared with SNe Type Ia
292 data in fig.5b. From the curves, we see that the theoretical
parametrized DE models in Galileon gravity is in agreement with
the union2 sample data and the SNe Type Ia 292 data.

\section{Analysis with Supernovae Type Ia 292 data (from Riess et al.(2004, 2007) and Astier et al.(2006))}
Here we have analyzed our theoretical parametrized DE models in
the light of the Supernovae Type Ia 292 data \cite{Riess1, Riess2,
Astier1}. The compiled data can be found in ref \cite{Chayanb}.
Bounds for the model parameters $(\omega_{0}, \omega_{1})$ have
been obtained by constraining the theoretical DE models with the
SN Type Ia 292 data, after fixing the cosmological parameters in
their viable range. Given below is the $\chi^{2}$ statistic which
has been used as a minimization technique for our data analysis,
\begin{equation}
\chi^{2}_{(SNe Type Ia)}=\Sigma
\frac{\left(H(z)-H_{obs}(z)\right)^{2}}{\sigma^{2}(z)}
\end{equation}
where $H_{obs}(z)$, $\sigma(z)$ are obtained from the SNe Type Ia
292 data table and $H(z)$ is the calculated value of the Hubble
parameter obtained from our theoretical model. The probability
distribution function is given as,
\begin{equation}
L=\int e^{-\frac{1}{2}\chi^{2}_{(SNe Type Ia)}}P(H_{0})~ dH_{0}
\end{equation}
where the prior distribution function for $H_{0}$ is $P(H_{0})$.
The best fit values of the model parameters $(\omega_{0},
\omega{1})$ obtained for different theoretical models by the
$\chi^{2}$ minimization technique is presented in table 5 along
with their respective minimum $\chi^{2}$ values. Contours, which
are showing the numerical range of the parameters, are obtained
for different confidence intervals and are given in the figs. 4a,
4b, 4c and 4d.

\vspace{1mm}
\begin{figure}[h]
~~~~~~~~~~~~~~~~~~~\includegraphics[height=3in]{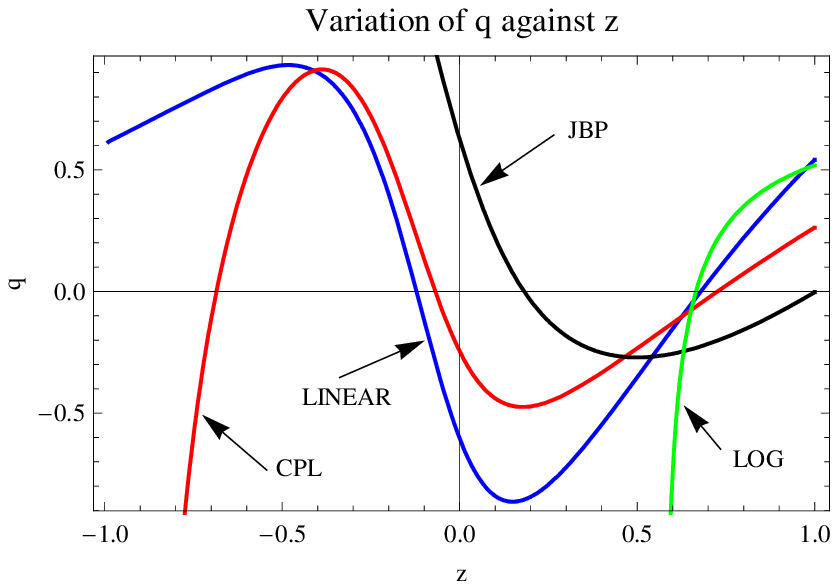}~~~~~~~~~\\
\vspace{1mm}
~~~~~~~~~~~~~~~~~~~~~~~~~~~~~~~~~~~~~~~~~~~~~~~~~~~~Fig.6~~~~~~~~~~~~~~~~~~~~~~~~~~~~~~~~\\
\vspace{2mm} \textit{\textbf{Fig 6} shows the variation of
deceleration parameter $q$ vs redshift $z$ for linear (Blue), CPL
(Red), JBP (Black) and logarithmic (Green) parametrizations.}
\end{figure}

\section{Discussions and conclusions}
In this assignment, we have proposed the FRW universe filled with
DM (perfect fluid with negligible pressure) along with dark energy
in the background of Galileon gravity. For describing DE
components, four parametrizations have been proposed for the
variations of EOS parameter $\omega(z)$. They are Linear, CPL, JBP
and Logarithmic parameterizations. We have presented the Hubble
parameter $H$ in terms of the observable parameters $\Omega_{m0}$,
$\Omega_{de0}$, $H_{0}$ with the redshift $z$ and the other
parameters like $\omega_0$ and $\omega_1$. We have chosen the
observed values of $\Omega_{m0} = 0.28$, $\Omega_{de0} = 0.72$ and
$H_{0} = 72 Kms^{-1} Mpc^{-1}$. From STERN data set (12 points),
we have obtained the best fit values of the arbitrary parameters
$\omega_{0}$ and $\omega_{1}$ (Table 2) by minimizing the
$\chi^{2}$ test and by fixing the other parameters. Similarly, the
best fit values of $\omega_{0}$ and $\omega_{1}$ are presented in
Table 3 and Table 4 for STERN+BAO and STERN+BAO+CMB respectively.
In Table 5, the best fit values of the parameters are given using
the Supernovae Type Ia 292 dataset. Likelihood contours have been
drawn for different confidence levels $(66\%, 90\%, 99\%)$
constraining the parameters $(\omega_{0}, \omega_{1})$. In the
figs. 1a, 1b, 1c and 1d contours have been drawn for STERN data.
In figs. 2a, 2b, 2c and 2d contours are drawn for STERN+BAO joint
data analysis. STERN+BAO+CMB joint data have been used to generate
the plots 3a, 3b, 3c and 3d. Contours for Supernovae Type Ia 292
data can be found in figs. 4a, 4b, 4c and 4d. Finally in figs. 5a
and 5b, using the best fit values for our different theoretical
parametrization models, $\mu$ vs redshift $(z)$ curves are plotted
and compared with Union compilation data and SNe Type Ia 292 data
respectively.

Figs. 1a to 1d have been generated for Stern data. In fig. 1a,
which is the plot for linear model, for $66\%$ confidence interval
$\omega_{0}$ and $\omega_{1}$ lies between $-2.39$ to $-1.51$ and
$0.78$ to $1.16$ respectively. Similarly for $90\%$ confidence
limit, the ranges for $\omega_{0}$ are $-2.52$ to $-1.58$ and
those for $\omega_{1}$ are $0.74$ to $1.18$. For $99\%$ confidence
limits the corresponding ranges for $\omega_{0}$ are $-2.67$ to
$-1.48$ and for $\omega_{1}$ are $0.68$ to $1.22$. From fig. 1b
for CPL parametrization we see that for $66\%$ confidence limits
$\omega_{0}$ lies between $-2.12$ to $-1.28$ and $\omega_{1}$ lies
between $0.76$ to $1.12$. For $90\%$ confidence limit $\omega_{0}$
lies between $-2.19$ to $-1.21$ and $\omega_{1}$ lies between
$0.72$ and $1.14$. Similarly for $99\%$ limits the ranges are
$-2.34$ to $-1.10$ for $\omega_{0}$ and $0.66$ to $1.18$ for
$\omega_{1}$. In fig. 1c, for JBP model, the corresponding ranges
for $\omega_{0}$ and $\omega_{1}$ are $-1.35$ to $-0.65$ and
$-2.20$ to $-1.45$ respectively, in $66\%$ confidence limits. For
$90\%$ confidence limits the ranges are $-1.50$ to $-0.60$ for
$\omega_{0}$ and $-2.22$ to $-1.38$ for $\omega_{1}$. Finally for
$99\%$ confidence interval the values are $-1.65$ to $-0.55$ for
$\omega_{0}$ and $-2.23$ to $-1.27$ for $\omega_{1}$.

In fig. 2a for linear model (joint analysis with Stern+BAO), the
ranges of parameters are $-11.50$ to $0.50$ for $\omega_{0}$ and
$0.12$ to $2.70$ for $\omega_{1}$ in $66\%$ confidence interval.
For $90\%$ confidence level, $\omega_{0}$ varies from $-12.30$ to
$0.61$ and $\omega_{1}$ varies from $-0.20$ to $2.80$. Finally for
$99\%$ confidence range, $\omega_{0}$ varies from $-13.20$ to
$0.90$ and $\omega_{1}$ varies from $-0.25$ to $2.85$. In fig. 2b,
for the CPL model, $\omega_{0}$ varies from $-2.86$ to $-1.89$ and
$\omega_{1}$ varies from $1.47$ to $2.25$ in $66\%$ confidence
interval. In $90\%$, confidence limits, $\omega_{0}$ varies from
$-2.96$ to $-1.80$ and $\omega_{1}$ varies from $1.42$ to $2.70$.
For $99\%$ confidence limits the variation is from $-3.08$ to
$-1.68$ for $\omega_{0}$ and $1.30$ to $2.11$ for $\omega_{1}$.
Plot for JBP is generated in fig. 2c. We see that in $66\%$
confidence interval, $\omega_{0}$ varies from $-1.31$ to $-0.69$
and $\omega_{1}$ varies from $-3.61$ to $-2.59$. In $90\%$
confidence limits, $\omega_{0}$ varies from $-1.39$ to $-0.61$ and
$\omega_{1}$ varies from $-3.72$ to $-2.48$. Similarly in $99\%$
confidence limits, $\omega_{0}$ and $\omega_{1}$ varies
respectively from $-1.49$ to $-0.56$ and $-3.82$ to $-2.30$.

From figs. 3a to 3d, plots are generated for Stern+BAO+CMB joint
data analysis. In fig. 3a, which is for linear model, the
acceptable range for $\omega_{0}$ is obtained as $-3.59$ to
$-2.52$ and that for $\omega_{1}$ is from $1.30$ to $1.73$ at
$66\%$ confidence level. Similarly at $90\%$ confidence level the
values of $\omega_{0}$ vary from $-3.76$ to $-2.42$ and for
$\omega_{1}$ they vary from $1.26$ to $1.76$. Finally for $99\%$
confidence interval, $\omega_{0}$ is from $-3.96$ to $-2.32$ and
$\omega_{1}$ is from $1.17$ to $1.81$. In fig. 3b, for CPL model,
$\omega_{0}$ varies from $-4.95$ to $-3.05$ and $\omega_{1}$
varies from $0.61$ to $0.81$ in $66\%$ confidence level. In $90\%$
confidence limits the ranges of $\omega_{0}$ and $\omega_{1}$ are
respectively from $-5.45$ to $-2.95$ and $0.55$ to $0.85$.
Similarly for $99\%$ limits, the acceptable range is from $-6.05$
to $-2.82$ for $\omega_{0}$ and from $0.49$ to $0.89$ for
$\omega_{1}$. In fig. 3c, the plot has been generated for JBP
model. For $66\%$ confidence interval, $\omega_{0}$ takes the
values from $-4.8$ to $-2.0$ and $\omega_{1}$ takes the values
from $-1.4$ to $-0.45$. At $90\%$ confidence interval, the
acceptable ranges for $\omega_{0}$ and $\omega_{1}$ are $-5.4$ to
$-1.9$ and $-1.50$ to $-0.43$ respectively. Finally at $99\%$
confidence limits $\omega_{0}$ varies from $-6.5$ to $-1.8$ and
$\omega_{1}$ varies from $-1.7$ to $-0.25$.

Plots for SNe Type Ia 292 data are generated in figs. 4a to 4d for
different parametrization models. In fig. 4a, for linear model, at
$66\%$ confidence level, $\omega_{0}$ varies from $-2.7142$ to
$-2.6980$ and $\omega_{1}$ changes from $0.5078$ to $0.5116$. At
$90\%$ confidence limits $\omega_{0}$ ranges from $-2.7159$ to
$-2.6964$ and $\omega_{1}$ ranges from $0.5074$ to $0.5118$.
Finally at $99\%$ confidence interval, the acceptable ranges are
$-2.7165$ to $-2.6945$ for $\omega_{0}$ and $0.5070$ to $0.5123$
for $\omega_{1}$. Fig. 4b is generated for the CPL model. Here at
$66\%$ confidence interval, $\omega_{0}$ ranges from $-2.6425$ to
$-2.6270$ and $\omega_{1}$ ranges from $0.5070$ to $0.5105$. At
$90\%$ confidence limits, $-2.6442$ to $-2.6247$ is the range for
$\omega_{0}$ and $0.5066$ to $0.5111$ is the range for
$\omega_{1}$. Similarly at $99\%$ confidence interval, the
acceptable ranges vary from $-2.6460$ to $-2.6239$ for
$\omega_{0}$ and $0.5062$ to $0.5116$ for $\omega_{1}$. Plot for
the JBP model has been obtained in fig. 4c. Here we see that,
$\omega_{0}$ and $\omega_{1}$ respectively range from $-2.3460$ to
$-2.3318$ and $-1.0072$ to $-0.9999$ for $66\%$ confidence limit.
At $90\%$ confidence limits $\omega_{0}$ attains values from
$-2.3475$ to $-2.3302$. Similarly the acceptable range for
$\omega_{1}$ is from $-1.0078$ to $-0.9990$. Likewise, at $99\%$
confidence limits the $\omega_{0}$ lies in the range $-2.3490$ to
$-2.3287$ and $\omega_{1}$ lies in the range $-1.0088$ to
$-0.9982$.

For all the four cases the Logarithmic parametrization is studied
in the figures 1d, 2d, 3d and 4d. In all the four figures it can
be seen that the likelihood contours are parallel lines for all
the three confidence levels. Unavailability of closed contours
hinders our analysis considerably and due to this unbounded nature
we are not able to put any finite bounds on both $\omega_{0}$ and
$\omega_{1}$ from the present study and with the given choice of
parameters. Looking at the figures, mathematically, one can
obviously argue that the bound for the parameters are $(-\infty,
\infty)$, but we do not think that it is physically acceptable.
So, may be we have to devise some other methods to constrain the
parameters for the logarithmic model, which we keep as an open
issue for the time being.

On a general note, it is observed that for Linear and CPL model,
with the increase in the value of $\omega_{0}$, the value of
$\omega_{1}$ decreases. But in case of JBP model, an exactly
opposite nature is witnessed. Increase in $\omega_{0}$ is
accompanied by an increase in $\omega_{1}$. It must also be noted
that for both the Linear and CPL model $\omega_{0}$ lies in the
negative range, whereas $\omega_{1}$ takes positive values. For
JBP model, the scenario is a bit different. Both $\omega_{0}$ and
$\omega_{1}$ lies in the negative range. These results and the
trends obtained for the involved parameters are quite different
from the ones obtained by Biswas et al. in \cite{Biswas1} using
Loop quantum cosmology as background gravity. In \cite{Biswas1},
it was found that JBP model allows a wider range of values for the
parameters compared to the linear and CPL models. Moreover with
the addition of BAO peak analysis and CMB term the reduces the
lower and upper freedom of $\omega_{0}$ and $\omega_{1}$. But in
the present study no such trend is observed.

In the fig. 5a, distance modulus $\mu(z)$ is plotted against
redshift parameter $z$ for all the four parametrization models and
then compared with Union2 data Sample. The data sample is
represented by a scatter diagram and the trajectories represent
the $\mu(z)$ vs $z$ plots for various models. It is seen that
barring the Logarithmic model, all the other models roughly fits
the dataset. At lower redshifts $(z<0.3)$ it is seen that the fit
is not quite as good as that at higher redshifts. The best fit is
obtained around $z=0.4$. Moreover for the Logarithmic model, there
is almost no fit with the data. In fig. 5b, the trajectories are
compared with SNe Type Ia 292 data. Here also it is seen that the
logarithmic model does not fit the dataset whereas the other three
models gives satisfactory fit with the data. In this case, the fit
is considerably improved at lower redshifts compared to the union
2 data sample. At higher redshifts the curve fitting is proper.
The best fit is obtained around $z=0.5$. Although the best fit
point has shifted to a higher redshift in case of SNe Type Ia 292
dataset compared to the Union2 dataset, yet the Sne data must be
preferred compared to its counterpart because it gives a better
fit with the parametrization models at lower redshifts with the
given choice of parameters. Alberto et al. in \cite{Alberto1} used
Bayesian analysis to predict that JBP parametrization is preferred
to CPL. But in this assignment, we did not find any result, which
will either support or criticize the prediction. Finally, it must
be stated that from the present study, we find that the predicted
theoretical Logarithmic model does not fit either of the
observational data. Moreover, for the logarithmic model, no finite
bounds for the parameters could be obtained.

In fig. 6, we have plotted the deceleration parameter $q$ vs the
redshift parameter $z$. Recently there have been quite a few works
regarding cosmological deceleration acceleration transition
\cite{Farooq1, Farooq2}. In these works it can be seen that the
transition from deceleration to acceleration occurs at $z=0.74\pm
0.05$. We have obtained the expressions for the deceleration
parameter for all the parametrization models using their best fit
$H(z)-z$ values of the parameters and then plotted them against
redshift in fig. 6. From the figure it is seen that the transition
from deceleration to acceleration regime occurs at $z=0.68$,
$z=0.74$, $z=0.98$ and $z=0.67$ for the linear, CPL, JBP and
logarithmic models respectively. So the result for the CPL model
is in complete accordance with the works of Farooq et al.
\cite{Farooq1, Farooq2}. Moreover the the values for the linear
and logarithmic models almost lie in the redshift spectrum of
transition obtained in Farooq et al. Only in case of the JBP model
the transition is early which contradicts the result obtained by
Farooq et al. Finally it must be pointed out that since we have
fixed the free parameters with particular values, we have lost the
overall generality of the study upto a certain extent.
Correspondingly the EoS contours obtained could have been much
broader, if these particular choices could have been avoided. So
the present study might have lost some of its numerical accuracy.
Nevertheless its cosmological significance is unquestionable.

\section*{Acknowledgements}

The authors sincerely acknowledge the facilities provided by the
Inter-University Centre for Astronomy and Astrophysics (IUCAA),
pune, India where a part of the work was carried out. The Authors
acknowledge the anonymous referee for enlightening comments that
helped to improve the quality of the manuscript.\\


\begin{thebibliography}{99}
\bibitem{Perlmutter} S. J. Perlmutter et al. :- {\it Nature} {\bf 391} 51 (1998).
\bibitem{Riess} A. G. Riess et al. [Supernova Search Team Collaboration] :- {\it Astron. J.} {\bf 116} 1009 (1998).
\bibitem{Riess1}  A. G. Riess et al. :- {\it Astrophys. J.} {\bf 607} 665 (2004).
\bibitem{Bennet} C. Bennet et al. :- {\it Phys. Rev. Lett.} {\bf 85} 2236 (2000).
\bibitem{Sperge} D. N. Spergel et al. :- {\it Astrophys. J. Suppl. Ser.} {\bf 170} 377 (2007).
\bibitem{Adel}  J. K. Adelman-McCarthy et al. :-  {\it Astrophys. J. Suppl. Ser.} {\bf 175} 297 (2008).
\bibitem{Kobayashi1} T. Kobayashi, H. Tashiro, D. Suzuki :- {\it Phys. Rev. D.} {\bf 81} 063513 (2010)
\bibitem{Eisenstein} D. J. Eisenstein et al. [SDSS Collaboration] :-  {\it Astrophys. J.} {\bf 633} 560 (2005).
\bibitem{Briddle} S. Briddle et al. :-  {\it Science} {\bf 299} 1532 (2003).
\bibitem{Spergel} D. N. Spergel et al. :-  {\it Astrophys. J. Suppl.} {\bf 148}, 175 (2003).
\bibitem{Peebles} P. J. E. Peebles, B. Ratra  :- {\it Astrophys. J.} {\bf 325} L17 (1988).
\bibitem{Cald} R. R. Caldwell, R. Dave, P. J. Steinhardt  :- {\it Phys. Rev. Lett.} {\bf 80} 1582 (1998).
\bibitem{Arme} C. Armendariz-Picon, V. F. Mukhanov, P. J. Steinhardt :- {\it Phys. Rev. Lett.} {\bf 85} 4438 (2000).
\bibitem{Sen}  A. Sen  :- {\it JHEP} {\bf 0207} 065 (2002).
\bibitem{Cald1}  R. R. Caldwell :-  {\it Phys. Lett. B} {\bf 545} 23 (2002).
\bibitem{Feng} B. Feng,  X. L. Wang,  X. M. Zhang :- {\it Phys. Lett. B} {\bf 607} 35 (2005).
\bibitem{Kamen} A. Y. Kamenshchik, U. Moschella, V. Pasquier :-  {\it Phys. Lett. B} {\bf 511} 265 (2001).
\bibitem{Debnath} U. Debnath, A. Banerjee, S. Chakraborty :-  {\it Class. Quantum Grav.} {\bf 21} 5609 (2001).
\bibitem{Cohen}  A. Cohen, D. Kaplan, A. Nelson :-  {\it Phys. Rev. Lett.} {\bf 82}, 4971 (1999).
\bibitem{Sahni}  V. Sahni, Y. Shtanov :-  {\it JCAP} {\bf 0311} 014 (2003).
\bibitem{Cai}  R. G. Cai :-  {\it Phys. Lett. B} {\bf 657} 228 (2007).
\bibitem{Wei} H. Wei,  R. G. Cai :- {\it Phys. Lett. B} {\bf 660} 113 (2008).
\bibitem{Paddy1} T. R. Choudhury,  T. Padmanabhan :- {\it Astron. Astrophys.} {\bf 429} 807 (2007).
\bibitem{Tonry} J. L. Tonry et al. :-  {\it Astrophys. J.} {\bf 594} 1 (2003).
\bibitem{Barris} B. J. Barris et al. :- {\it Astrophys. J.} {\bf 602} 571 (2004).
\bibitem{Lu} J. Lu et al. :- {\it Phys. Lett. B} {\bf 662} 87 (2008).
\bibitem{Jun}  L. Dao-Jun, L. Xin-Zhou :- {\it Chin. Phys. Lett.} {\bf 22} 1600 (2005).
\bibitem{Dvali} G. R. Dvali, G. Gabadadze, M. Porrati :- {\it Phys. Lett. B} {\bf 484} 112 (2000).
\bibitem{An}  A. De Felice,  T. Tsujikawa :-  {\it arXiv}: 1002.4928 [gr-qc].
\bibitem{Noj0} S. Nojiri, S. D. Odintsov :- {\it arXiv}: 1011.0544 [gr-q].
\bibitem{clif}  T. Clifton, J. Barrow :- {\it Phys. Rev. D} {\bf 72} 103005 (2005).
\bibitem{Yer} K. K. Yerzhanov et al. :- {\it arXiv}:1006.3879v1 [gr-qc] (2010).
\bibitem{Noj} S. Nojiri, S. D. Odintsov :- {\it Phys. Lett. B} {\bf 631} 1 (2005).
\bibitem{An1} I. Antoniadis, J. Rizos,  K. Tamvakis :- {\it Nucl. Phys. B} {\bf 415} 497 (1994).
\bibitem{Hora}  P. Horava :- {\it JHEP} {\bf 0903} 020 (2009).
\bibitem{Brans} C. Brans, H. Dicke :- {\it Phys. Rev.} {\bf 124} 925 (1961).
\bibitem{Nicolis1} A. Nicolis,  R. Rattazzi, E. Trincherini :-  {\it Phys. Rev. D} {\bf 79} 064036(2009)
\bibitem{Melchiorri1} A. Melchiorri, L. Mersini, M. Trodden :- {\it Phys. Rev. D} {\bf 68} 043509(2003)
\bibitem{Seljak1} U. Seljak, A. Slosar, P. Mcdonald :- {\it JCAP} {\bf 0610} 014 (2006)
\bibitem{Paddy2} T. Padmanabhan, T. R. Chowdhury :- {\it Mon. Not. R. Astron. Soc.} {\bf 344} 823 (2003)
\bibitem{Amanullah1} R. Amanullah et al. :- {\it Astrophys. J.} {\bf 716} 712 (2010).
\bibitem{Ranjit1} C. Ranjit, P. Rudra,  S. Kundu :- {\it Astrophys. Space Sci.} {\bf 347} 423(2013)
\bibitem{Ranjit2} C. Ranjit, P. Rudra, U. Debnath :- {\it Can. J. Phys.} {\bf 92} 1667 (2014)
\bibitem{Chakraborty1} S. Chakraborty, U. Debnath, C. Ranjit :- {\it Eur. Phys. J. C.} {\bf 72} 2101(2012)
\bibitem{Paul1} P. Thakur, S. Ghose, B. C. Paul:-  {\it Mon. Not. R. Astron. Soc.} {\bf 397} 1935 (2009)
\bibitem{Paul2} B. C. Paul, S. Ghose, P. Thakur :- {\it Mon. Not. R. Astron. Soc.} {\bf 413} 686 (2011)
\bibitem{Silva1} F. P. Silva, K. Koyama :-  {\it Phys. Rev. D} {\bf 80} 121301 (2009)
\bibitem{Deffayet1} C. Deffayet, G. Esposito-Farese, A. Vikman :-  {\it Phys. Rev. D} {\bf 79} 084003 (2009)
\bibitem{Deffayet2} C. Deffayet, S. Deser,  G. Esposito-Farese :-  {\it Phys. Rev. D} {\bf 80} 064015 (2009)
\bibitem{Chow1} N. Chow, J. Khoury :-  {\it Phys. Rev. D} {\bf 80} 024037 (2009)
\bibitem{Cooray1}A. R. Cooray, D. Huterer :- {\it Astrophys. J.} {\bf 513} L95 (1999)
\bibitem{Chevallier1} M. Chevallier, D. Polarski :- {\it Int. J. Mod. Phys. D.} {\bf 10} 213 (2001)
\bibitem{Linder1} E. V. Linder :- {\it Phys. Rev. Lett.} {\bf 90} 091301 (2003)
\bibitem{Brax1} P. Brax, J. Martin :- {\it Phys. Lett. B.} {\bf 468} 40 (1999)
\bibitem{Jassal1}H. K. Jassal, J. S. Bagla and T. Padmanabhan :- {\it Mon. Not. R. Astron. Soc.} {\bf 356} L11 (2005)
\bibitem{Efs1}G. Efstathiou :- {\it Mon. Not. R. Astron. Soc.} {\bf 310} 842 (1999)
\bibitem{Silva2}R. Silva, J. S. Alcaniz, J. A. S. Lima :- {\it Int. J. Mod. Phys. D.} {\bf 16}469 (2007)
\bibitem{Stern1} D. Stern et al. :-  {\it JCAP} {\bf 1002} 008 (2010).
\bibitem{Bond1} J. R. Bond et al. :- {\it Mon. Not. Roy. Astron. Soc.} {\bf 291} L33 (1997)
\bibitem{Efstathiou1} G. Efstathiou, J. R. Bond :- {\it Mon. Not. R. Astro. Soc.} {\bf 304} 75 (1999)
\bibitem{Nessaeris1} S. Nessaeris,  L. Perivolaropoulos :- {\it JCAP} {\bf 0701} 018 (2007).
\bibitem{Komatsu1} E. Komatsu et al. :- {\it Astrophys. J. Suppl.} {\bf 192} 18 (2011).
\bibitem{Perlmutter1} S. J. Perlmutter et al. :- {\it Astrophys. J.} {\bf 517} 565 (1999).
\bibitem{Riess2} A. G. Riess et al. :-  {\it Astrophys. J.} {\bf 659} 98 (2007).
\bibitem{Kowalaski1} M. Kowalski et al. :- {\it Astrophys. J.} {\bf 686} 749 (2008).
\bibitem{Astier1} P. Astier et. al. :- {\it Astron. Astrophys.} {\bf 447} 31 (2006)
\bibitem{Chayanb} C. Ranjit, U. Debnath :- {\it Astrophys. Space Sci.} {\bf 354} 2126 (2014), {\it arXiv:1409.7057 [phys-gen.ph]}
\bibitem{Biswas1} R. Biswas, U. Debnath :- {\it Eur. Phys. J. C.} {\bf 73} 2424 (2013)
\bibitem{Alberto1} J. Alberto Vazquez, M. Bridges, M.P. Hobson, A.N. Lasenby :- {\it arXiv:1205.0847 [astro-ph.CO]}
\bibitem{Farooq1} O. Farooq, B. Ratra :- {\it arXiv}: 1301.5243 [astro-ph.CO] (2013).
\bibitem{Farooq2} O. Farooq, S. Crandall, B. Ratra :- {\it arXiv}: 1305.1957 [astro-ph.CO] (2013).
\end{thebibliography}
\end{document}